\documentclass[
reprint, twocolumn, secnumarabic%
,tightenlines%
,amssymb,
amsmath,
nobibnotes,
nofootinbib, aps, prl, showpacs,showkeys]{revtex4}

\usepackage{subfigure}
\usepackage{tikz}
\usepackage{docs}%
\usepackage{bm}%
\usepackage{graphicx} %
\usepackage{rotating} %
\usepackage{hyperref} %
\usepackage{url} %
\usepackage{srcltx} %
\usepackage{float}
\expandafter \ifx \csname package@font\endcsname\relax\else
\expandafter \expandafter \expandafter
\usepackage
\expandafter \expandafter
\expandafter{\csname package@font\endcsname}%
\fi
\usepackage{epstopdf}
\usepackage{slashed}
\usepackage{ulem}
\usepackage{srcltx}
\usepackage{soul}
\usepackage{amsmath}
\usepackage{natbib} 
\usepackage{ulem} 

\usepackage{float} 
\extrafloats{200}


\usepackage{epstopdf}

\usepackage{enumerate}

\begin{document}

\title{Model dependence of the polarization asymmetries in weak pion production off the nucleon}

\author{Krzysztof M. Graczyk}
\email{krzysztof.graczyk@uwr.edu.pl}

\author{Beata E. Kowal}

\affiliation{Institute of Theoretical Physics, University of Wroc\l aw, plac Maxa Borna 9,
50-204, Wroc\l aw, Poland}

\date{\today}%

\begin{abstract}

The work presents the studies of the polarization observables  in the single pion production (SPP) induced by the interaction of the muon neutrino (antineutrino) with nucleons. The SPP cross-sections and spin asymmetries are computed within two phenomenological models. One is basing on the nonlinear sigma model \cite{Hernandez:2007qq}; the other has origin in linear sigma model~\cite{Fogli:1979cz}.  Firstly, we show that the final nucleon polarization and target spin asymmetries are good observables to obtain information about the $C_5^A$ axial form factor.  Secondly, we demonstrate that the nucleon polarization and the target spin asymmetries are sensitive to the relative phase between resonance and nonresonance amplitudes. We conclude that the polarization of the final nucleon and the target spin asymmetry are promising observables for testing SPP models, including studies of the axial content of $\Delta(1232)$ resonance and unitarization procedures.   

\end{abstract}

\maketitle

\section{Introduction}

A single pion production (SPP) induced by the weak neutrino-nucleon interaction is an essential process for experimental studies of the neutrino oscillation phenomenon~\cite{Alvarez-Ruso:2014bla,Mosel_annurev-nucl-102115-044720,Katori:2016yel,Athar:1970esm,SajjadAthar:2020nvy,Alvarez-Ruso:2020ezu}. In the long and short baseline experiments~\cite{AguilarArevalo:2007it,Abe:2019vii,Evans:2013pka,Ayres:2004js}, the oscillation parameters are obtained from the analysis of the quasielastic (QE) neutrino-nucleus events. Nevertheless, the SPP events contribute to the background for measuring the QE events and the background for the measurement of the electrons produced in charged current $\nu_e$-nucleus scattering where $\nu_e$ is the result of the $\nu_\mu \to \nu_e$ oscillation.

The neutrino oscillation phenomenon has been studied for about five decades~\cite{Bilenky:1978nj,Bilenky:2012qb}. Recently the CP-violation phase has been measured by Tokai-to-Kamioka (T2K) collaboration~\cite{Abe:2019vii}.  The next step, in exploring the neutrino properties, is to verify whenever neutrinos oscillate differently from antineutrinos.  
\begin{figure}
    \centering
    \includegraphics{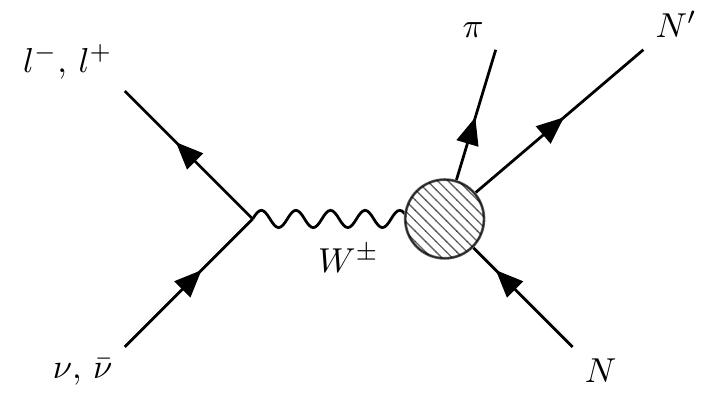}
    \caption{Schematic diagram of the weak SPP  in the charged current neutrino (antineutrino) scattering off nucleon.}
    \label{fig:SPP_simple_diagram}
\end{figure}

This paper considers the SPP induced by the charged-current (CC) $\nu_\mu$-nucleon interaction in the energy range characteristic for the experiments with accelerator source of neutrinos such as T2K. Therefore we shall consider the neutrino energy of the order of $1$~GeV.
Schematically the SPP process, see Fig.~\ref{fig:SPP_simple_diagram},  can be written as
\begin{equation}
    \nu_\mu + N  \to \mu^- + {N'} + \pi,
\end{equation}
where $N$ and $N'$ denote target and recoil nucleon, respectively.

In the discussed energy range the dominant contribution to the SPP scattering amplitude comes from the weak nucleon-$\Delta(1232)$ excitation. The $\Delta$ resonance decays into the pion-nucleon system.  However, the completeness of the description requires the inclusion of contribution from the nonresonance production of single pions.
     
The theoretical models for the weak pion production have been developed for more than sixty years~\cite{LlewellynSmith:1971uhs}. However, it is important to underline that there is no unique method to describe the SPP processes.  Indeed, there are various methods and strategies~\cite{Adler:1968tw,Schreiner:1973mj,Ravndal:1973xx,Fogli:1979cz,Gershtein:1980vd,Rein:1980wg,Rein:1987cb,AlvarezRuso:1997jr,Sato:2003rq,Kuzmin:2004ya,Lalakulich:2006sw,Hernandez:2006yg,Graczyk:2007bc,Graczyk:2007bc,Berger:2007rq,Barbero:2008zza,Graczyk:2008zz,Leitner:2008ue,Graczyk:2009qm,Lalakulich:2010ss,Mariano:2011zz,Sobczyk.:2012zj,Serot:2012rd,Graczyk:2014dpa,Alam:2015gaa,Alvarez-Ruso:2015eva,Gonzalez-Jimenez:2016qqq,Hernandez:2016yfb,Yao:2018pzc,Nakamura:2015rta,Nakamura:2018ntd,Sobczyk:2018ghy,Yao:2019avf,Rocco:2019gfb}. The theoretical predictions usually are biased by a choice of the model parameters and the way the model is constructed. 

A typical strategy in modeling SPP is to fine-tune the model parameters basing on the analysis of two bubble chamber experiments: Argonne National Laboratory (ANL)~\cite{Radecky:1981fn} and Brookhaven National Laboratory (BNL)~\cite{Kitagaki:1986ct}. In both projects, neutrino-deuteron scattering was investigated. Since modeling of the nuclear effects for the deuteron target is a relatively simple task~\cite{Singh:1986xh}, therefore, the analysis of these data allows to obtain the information about the elementary neutrino-nucleon vertex, see, e.g., \cite{Graczyk:2014dpa}. 

In the last years, several analyses of ANL and BNL data have been performed. It has been shown that the ANL and BNL data are statistically consistent~\cite{Graczyk:2009qm,Hernandez:2010bx,Wilkinson:2014yfa}.    Although both the ANL and BNL experiments provided several distributions of measured events, these data are not informative enough to constrain and verify the SPP models thoroughly.

The complete information about the weak neutrino-nucleon scattering amplitudes is encoded in the polarization observables. Therefore, the polarization properties in the $\nu$N scattering have been investigated by many groups. However, most of the studies are dedicated to the final particles' polarization properties in the quasielastic and elastic $\nu N$ scattering~\cite{Adler1963,PhysRev.168.1662,Pais:1971er,Block:1965zol,LlewellynSmith:1971uhs, PhysRev.126.2239,PhysRevD.3.733,TARRACH197470,PhysRevD.10.993,PhysRevD.18.123,PhysRevD.28.2875,PhysRevD.32.2921,Jachowicz:2004we,PhysRevC.71.034604,Hagiwara:2003di,Kuzmin:2004yb,Kuzmin:2003ji,Graczyk:2004vg,Graczyk:2004uy,Levy:2004rk,PhysRevD.69.114019,Lagoda:2007zz,Valverde:2006yi,PhysRevC.77.034606,Bilenky:2013iua,Bilenky:2013fra,Graczyk:2017ngi,Fatima:2018tzs,Sobczyk:2019urm,Graczyk:2019xwg,Graczyk:2019opm,Fatima:2020pvv,Tomalak:2020zlv,Luis_polarization}. 
In Refs. \cite{HAGIWARA2005140,Kuzmin:2003ji} the polarization properties of the final lepton produced in
the weak SPP induced by $\nu N$  interaction has been studied as well. Moreover, quite recently, we have studied the dependence of the polarization observables on resonance and nonresonant contributions in SPP processes. Using two different models for the SPP,  HNV (Hernandez, Nieves, Valverede)~\cite{Hernandez:2007qq}, and FN (Fogli, Narduli) \cite{Fogli:1979cz} in Ref. \cite{Graczyk:2017rti} we made predictions for the polarization of the final nucleon and the charged lepton. In the following paper \cite{Graczyk:2019blt} we studied the target spin asymmetry. We showed that the polarization observables contain unique information about resonance and nonresonant background contributions, and they can be used to constrain the SPP models.

In this paper, we continue our previous investigations. In addition, we investigate the dependence of the lepton and the final nucleon polarization as well as target spin asymmetry on the SPP model inputs such as $C_5^A$ axial form factor and relative phase between resonance and nonresonant amplitudes.   

The weak nucleon-$\Delta(1232)$ transition current has the vector-axial structure.  The  information about the vector part is hidden in three independent form factors $C_3^V$, $C_4^V$, and $C_5^V$. These functions are obtained from  the analysis of the electroproduction scattering data; see, e.g., \cite{Lalakulich:2006sw,Sato:2003rq,Graczyk:2007bc,Barbero:2008zza,Graczyk:2014dpa}. The axial current is described by four form factors, namely, $C_3^A$,  $C_4^A$, $C_5^A$, and $C_6^A$. But $C_5^A$ from factor is the most important. Indeed, usually it is assumed that  
\begin{eqnarray}
\label{Eq:Adler_relation_for_C3A_and_C4A}
C_3^A=0, \quad \mathrm{and} \quad C_4^A = - \frac{C_5^A}{4} .
\end{eqnarray}
The first constraint comes from simple quark model derivations (see, e.g., \cite{Graczyk:2007bc}) the other from the dispersion models \cite{Adler:1968tw}.  
The partially conserved axial current  (PCAC) hypothesis implies that $C_6^A \sim C_5^A$~\cite{Thomas_book}.  

The  $C_5^A$ is also computed theoretically  within chiral field theory \cite{Hemmert:1994ky,BarquillaCano:2007yk,Geng:2008bm,Procura:2008ze,Unal:2018ruo,Kucukarslan:2014bla} and the lattice QCD~\cite{Alexandrou:2009vqd,Alexandrou:2013opa} as well. But still the most reliable information about $C_5^A$ comes from the analysis of the $\nu $-deuteron ANL and BNL scattering data~\cite{Schreiner:1973mj,Liu:1995bu,Lalakulich:2005cs, AlvarezRuso:1998hi,Leitner:2008fg,Graczyk:2007bc,Hernandez:2007qq,Graczyk:2009qm,Hernandez:2010bx,Graczyk:2014dpa}. In this work we shall investigate the  impact of the $C_5^A$ axial form factor on the polarization observables.

In non-unitary models, such  as HNV or FN, the relative phase factor, denoted here by $\exp(i \Phi)$,  between resonance ($\mathcal{M}_R$) and nonresonant background ($\mathcal{M}_{NB}$) amplitudes  
\begin{equation}
\label{Eq:general_structure}
    \mathcal{M}_{tot} =  \mathcal{M}_{R} + \exp(i \Phi) \mathcal{M}_{NB}.
\end{equation}
is equal to one ($\Phi=0$). However, in principle, the relative phase does not have to vanish. Indeed the unitarization procedure, basing on the Watson theorem,  allows modifying the vector and axial N-$\Delta$ transition currents dynamically by introducing small phases \cite{Alvarez-Ruso:2015eva}.  In this work, we aim to check the sensitivity of the polarization observables on the relative phase. To achieve this goal, we shall discuss a toy model scenario in which we shall keep the relative phase factor between resonance and nonresonant amplitudes.

The paper is organised as follows: in Sec.~\ref{sec:Formalism} we introduce the necessary formalism. Sec.~\ref{sec:Results} contains numerical results, main conclusions and summary.
\begin{figure*}
    \centering
    \includegraphics[width=\textwidth]{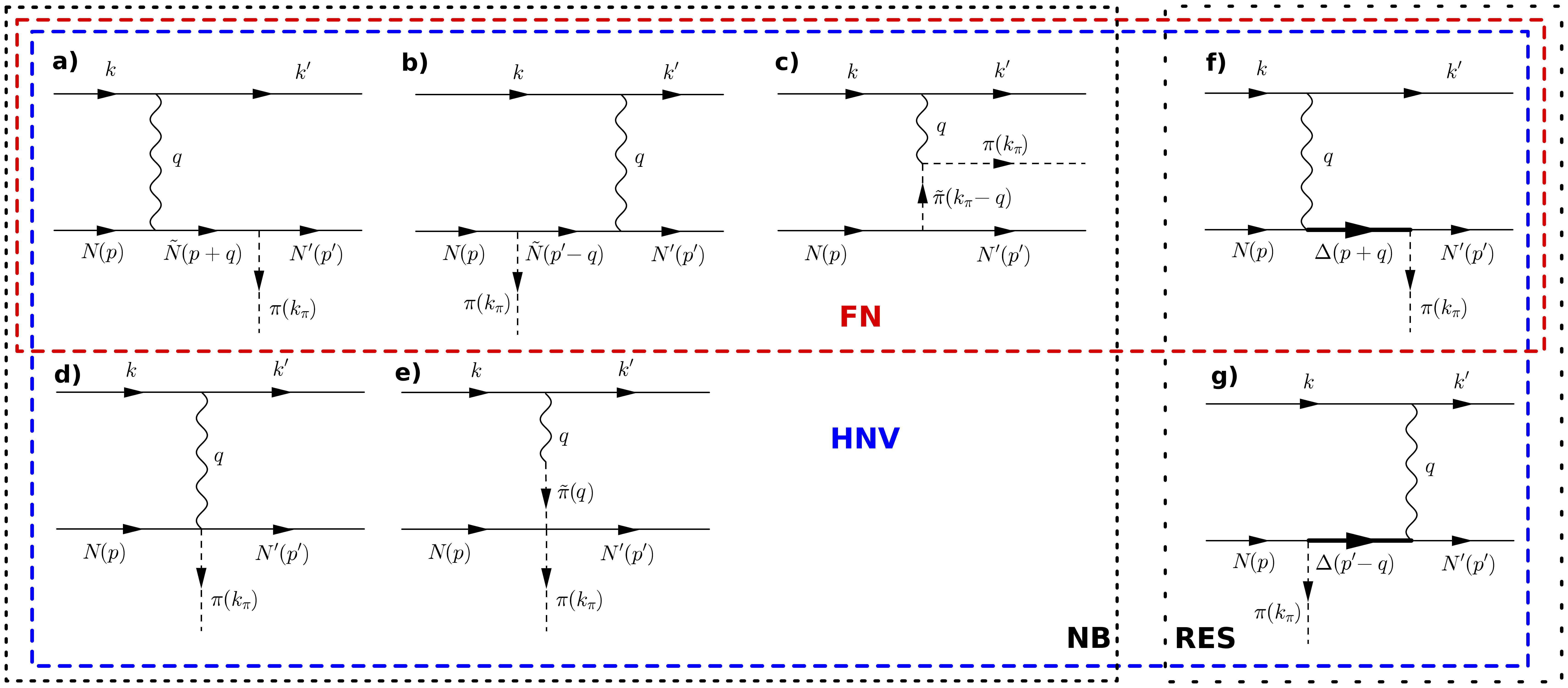}
    \caption{Feynman diagrams contributing to the FN (diagrams in red-dashed frame) and HNV (diagrams in blue-dashed frame) models, respectively. Resonance contribution is given by diagrams (f) and (g) whereas nonresonance contribution is represented by diagrams (a)-(e).  \label{fig:model-diagrams}}
\end{figure*}

\section{Formalism}
\label{sec:Formalism}


We consider three types of the polarization processes, namely:
\begin{itemize}
    \item neutrino (antineutrino\footnote{We explicitly write only the neutrino-nucleon scattering processes, but we consider also corresponding antineutrino-nucleon scattering processes as well. }) scatters on the nucleon, and the polarization of the charged lepton is measured:
    \begin{equation}
    \label{Eq:process1}
    \nu_\mu (k) + N(p) \to \vec{\mu}(k',s_{l}) + N'(p') + \pi(k_\pi); 
\end{equation}
\item neutrino scatters on the nucleon, and the polarization of the outgoing nucleon is measured:
\begin{equation}
    \label{Eq:process2}
    \nu_\mu (k) + N(p) \to \mu(k') + \vec{N'}(p',s_{N'}) + \pi(k_\pi); 
\end{equation}
\item neutrino scatters on the polarized target:
\begin{equation}
    \label{Eq:process3}
    \nu_\mu (k) + \vec{N}(p,s^N) \to \mu(k') + N'(p') + \pi(k_\pi),
\end{equation}
and the target spin asymmetry is measured.
\end{itemize}
In the above expressions  by $k$, $k'$, and $p$, $p'$ we denote the four-momenta of the neutrino, the muon, and the target nucleon, the final nucleon, respectively, whereas $s_l$, $s^N$, and $s^{N'}$ denote the spin vectors of the outgoing lepton, target nucleon, and recoil nucleon, respectively. We perform calculations in the laboratory frame, hence, in the last case 
\begin{equation}
s^{N \mu} = (0,\mathbf{s}^N).
\end{equation}

In principle the cross section depends on the spin four-vectors as it follows: 
	\begin{equation}
	\label{Eq:cross-section}
	{d\sigma} 
	= 
	{d\sigma_0}  \left(1 + \mathcal{P}_l^\mu s_{\mu}^l + \mathcal{T}_{N}^\mu s_{\mu}^N  + \mathcal{P}_{N'}^\mu s_{\mu}^{N'} 
	+ \dots \right).  
	\end{equation}
The asymmetries: $\mathcal{P}_l^\mu$ and $\mathcal{P}_{N'}^\mu$ are the polarization four-vectors of the muon and the final lepton, whereas  $\mathcal{T}_{N}^\mu$ is the target spin asymmetry.

Each spin polarization, as well as the target spin  asymmetry, depends on three independent components, namely:
\begin{eqnarray}		
	\mathcal{P}_l^{\mu}
	&=&
	\mathcal{P}^l_L  \zeta_L^{\mu}+\mathcal{P}^l_T \zeta_T^{\mu}+\mathcal{P}^l_N  \zeta_N^{\mu},
	\quad \\ 
	\mathcal{P}_{N'}^{\mu}
	&=& \mathcal{P}^{N'}_L  \xi_L^{\mu}+\mathcal{P}^{N'}_T \xi_T^{\mu}+\mathcal{P}^{N'}_N  \xi_N^{\mu},
\end{eqnarray}
where $\zeta_{L, T, N}^{\mu}$ and $\xi_{L,T,N}^{\mu}$ are the spin basis vectors normalized so that:
\begin{equation}
    -1 = \zeta_{L, T, N}^2 = \xi_{L,T,N}^2.
\end{equation}
Here, we consider the same spin vector basis as in Sec.~III of Ref.~\cite{Graczyk:2017rti}, see also Appendix~\ref{Appendix:Spin_baisis}.

Eventually the lepton and final nucleon polarizations are given by the ratios:
\begin{eqnarray}
\mathcal{P}_{X}^l &=& 
\frac{d \sigma(\zeta_X) - d \sigma(-\zeta_X)}{d \sigma(\zeta_X) + d \sigma(-\zeta_X)}
\\
\mathcal{P}_{X}^{N'} &=& 
\frac{d \sigma(\xi_{X}) - d \sigma(-\xi_X)}{d \sigma(\xi_X) + d \sigma(-\xi_X)},
\end{eqnarray}
where $X=L,T,N$.

In the last discussed process we compute 
the target asymmetry given by:
\begin{equation}		
			\mathcal{T}_N^{\mu}=\mathcal{T}^N_L  \chi_L^{\mu} + \mathcal{T}^N_P (\alpha) \chi_P^{\mu} (\alpha).
\end{equation}
where $\alpha \in (0,2\pi)$, $\chi_P^{\mu}$ is perpendicular to the neutrino velocity. We distinguish two components of $\chi_P^{\mu}$ normal to the scattering plane ($\alpha = 0 $) and transverse ($\alpha = 90^o $), which lays in the scattering plane spanned by $\mathbf{k}$ and $\mathbf{k'}$. Then 
\begin{equation} 
\chi_P^\mu (\phi)=\cos(\phi) \chi_N^\mu+\sin(\phi)\chi_T^\mu .
\end{equation}
Similarly as for polarization components we compute the asymmetries from the expression:
\begin{equation}
\mathcal{T}_{X}^N = 
\frac{d \sigma(\chi_X) - d \sigma(-\chi_X)}{d \sigma(\chi_X) + d \sigma(-\chi_X)}.
\end{equation}
In Ref. \cite{Graczyk:2019blt} we kept a bit different notation, namely, longitudinal component  $\mathcal{T}^N_L$ was denoted as $\mathcal{A}_\parallel$ and 
$\mathcal{T}^N_P$ as $\mathcal{A}_\perp$. We have change the notation to be consistent with our last work made for the quasielastic neutrino-nucleon scattering \cite{Graczyk:2019xwg}.  

For a more detailed description of the implementation of the asymmetries, see Sec.~II of Ref. \cite{Graczyk:2019blt}, and Appendix~\ref{Appendix:Spin_baisis}. 



As we have mentioned above, we consider two SPP models: HNV and FN. The first has its origin in the nonlinear sigma model, whereas the other bases on the linear sigma model. As a result, they differ in the treatment of the $\pi N N$ vertex. Indeed, in the first model, the vertex has a pseudovector structure, whereas, in the other, it has the pseudo-axial structure~\cite{deSwart:1997ep}.

In both descriptions, one can separate the resonance and nonresonant contributions.
Initially, the FN model contained four resonance amplitudes. One amplitude describes the $N-\Delta$ excitation, whereas the rest refer to the nucleon's excitation to the three states from the second resonance region. However, in the present discussion, we omit these contributions. 

The total scattering amplitude, $\mathcal{M}_{tot}$, is given by the contribution from: 
\begin{itemize}
    \item two resonance diagrams (f) and (g) and five nonresonant background diagrams (a)-(e), in the HNV model, see Fig.~\ref{fig:model-diagrams};
    \item one resonance diagram (f) and three nonresonance diagrams (a)-(c) in the FN model, see Fig.~\ref{fig:model-diagrams}. 
\end{itemize}
For a detailed description of our implementation, see Ref.~\cite{Graczyk:2017rti}.


In both discussed SPP approaches to model the weak nucleon-$\Delta$ transition, the $\Delta$ resonance is described by Rarita-Schwinger field~\cite{Rarita:1941mf}, denoted by $\Psi_\alpha$, and the hadronic transition current has the form:
\begin{equation}
J^\nu \sim    \Psi_\nu(p+q) {\Gamma^{\nu\mu}}(p,q) u(p),
\end{equation}
where $q$ denotes the four-momentum transfer
\begin{equation}
    q = k-k',
\end{equation}
and $q^\mu = (\omega,\mathbf{q})$; $u(p)$ is the Dirac spinor of incoming nucleon with momentum $p$. The vertex  $\Gamma^{\nu\mu}$   has the following vector-axial structure:
\begin{equation}
\label{Eq:Gamma_V_A}
{\Gamma^{\nu\mu}}(p,q)=\left[V^{\nu\mu}(p,q) + A^{\nu\mu}(p,q) \right]\gamma_5.
\end{equation}

The vector component of the transition current reads \cite{Jones:1972ky}: 
\begin{widetext}
\begin{eqnarray}
V^{\nu\mu}(p,q)&=&
\frac{C_3^V}{M}(g^{\nu\mu} \slashed{q} - q^{\nu}\gamma^{\mu})
+ 
\frac{C_4^V}{M^2}(g^{\nu\mu} q\cdot (p+q) - q^{\nu}({p}^{\mu}+q^\mu))
 +
\frac{C_5^V}{M^2}(g^{\nu\mu} q \cdot p - q^{\nu}p'^{\mu}),
\end{eqnarray}
\end{widetext}
where $M$ denotes the average nucleon mass. 

In the HNV model, we implement the vector form factors from Ref.~\cite{Graczyk:2014dpa}.
In the FN mode the vector structure is oversimplified, namely: 
\begin{equation}
\label{Eq:FN_vertex_vector_simplification}
C_5^V(q^2)=0,  
\end{equation}
where $q^2 \equiv q_\mu q^\mu$
as well as 
\begin{equation} 
C_4^V(q^2) = - \frac{M}{W} C_3^V(q^2),
\end{equation}
where $W^2 = (p+q)^2$. We see that for the FN model it is enough to know only $C_3^V$ form factor, see Appendix B of Ref. \cite{Graczyk:2017rti}.

The axial current reads~\cite{Schreiner:1973mj}: 
\begin{widetext}
\begin{eqnarray}
A^{\nu\mu}(p,q)&=&
\left( \frac{C_3^A}{M}(g^{\nu\mu} \slashed{q} - q^{\nu}\gamma^{\mu})
+
\frac{C_4^A}{M^2}(g^{\nu\mu} q\cdot (p+q) - q^{\nu}({p}^{\mu}+{q}^{\mu}))
+g^{\nu\mu}C_5^A
+ \frac{C_6^A}{M^2}q^{\nu}q^{\mu} 
\right)\gamma_5.
\end{eqnarray}
\end{widetext}

In our numerical experiments we  assume  that
\begin{equation}
\label{Eq:PCAC}
C_6^A(q^2) =   \frac{M^2}{m_\pi^2 - q^2} C_5^A(q^2),
\end{equation}
It is  supported  the partially conserved vector current (PCAC) hypothesis. Additionally in the HNV model  we consider the constraints given by Eq.~\ref{Eq:Adler_relation_for_C3A_and_C4A} whereas in the FN model
\begin{equation}
\label{Eq:CA_FN}
    C_4^A(q^2) = C_3^A(q^2) = 0.
\end{equation}

When one accepts the simplification given by Eq. \ref{Eq:Adler_relation_for_C3A_and_C4A} or Eq. \ref{Eq:CA_FN} then the $C_5^A$ axial form factor plays the most important role in modeling electroweak resonance contribution to the SPP cross-section. 

In this work, we assume  the dipole parametrization of $C_5^A$, namely,
\begin{equation}
 \label{Eq:C5A_dipole}
    C_5^A(q^2) = \frac{C_5^A(0)}{\displaystyle \left(1 - \frac{q^2}{M_A^2}\right)^2}, 
\end{equation}
where $M_A$ is the axial mass, a parameter obtained from the analysis of the neutrino-nucleon SPP scattering data. The parameter $C_5^A(0)$  can be
constrained by the  off-diagonal Goldberger-Treiman relation \cite{Thomas_book,BarquillaCano:2007yk}:
\begin{equation}
\label{C5A_0_current} C_5^A(0) = \frac{g_{\pi N\Delta}  f_\pi
}{\sqrt{6} M} =  1.15 \pm 0.01,
\end{equation}
where $g_{\pi N\Delta}(q^2 = m_\pi^2 )=28.6 \pm 0.3$ and $f_\pi =
92.4$~MeV. However, in many phenomenological investigations, for instance in Refs.~\cite{Hernandez:2007qq, BarquillaCano:2007yk, Hemmert:1994ky}, the value of $C_5^A(0)$ is treated as additional model parameter and it is extracted from the data analysis \cite{Barish:1978pj,Graczyk:2009qm, Hernandez:2007qq,Hernandez:2010bx,Graczyk:2014dpa}. Indeed in Ref.~\cite{Graczyk:2009qm} combined analysis of ANL and BNL data lead to $C_5^A(0) = 1.10_{-0.08}^{+0.09}$ and $M_A=1.10_{-0.14}^{+0.15}$~GeV.
In this paper, we treat $C_5^A(0)$ as the SPP model parameter.


As we have written in the introduction, the scattering amplitude is a sum of resonance and nonresonance amplitudes, see Eq. \ref{Eq:general_structure}. In order to complicate the picture, we introduce the relative phase factor between resonance and nonresonance parts. Then the square of the module of the scattering amplitude reads:
\begin{eqnarray}
 \label{Eq:old_amp}
 |\mathcal{M}_{tot}|^2 &=&   |\mathcal{M}_{R}|^2 +| \mathcal{M}_{NR} |^2 + 2 \mathrm{Re}\left(\mathcal{M}_{R} \mathcal{M}_{NR}^*\right)
    \label{Eq:M2_1}
    \\
    &    &
    -4 \left( \sin^2\frac{\Phi}{2} \mathrm{Re}\left(\mathcal{M}_{R} \mathcal{M}_{NR}^*\right) \right. \nonumber \\
    &    & \left.
    - \sin\frac{\Phi}{2}\cos\frac{\Phi}{2}\mathrm{Im}\left(\mathcal{M}_{R} \mathcal{M}_{NR}^*\right)\right).
    \label{Eq:new_amp}
\end{eqnarray}
In above we distinguish two contributions, namely,  (\ref{Eq:M2_1}) and (\ref{Eq:new_amp}). The expression (\ref{Eq:M2_1}) describes  the case with $\Phi =0$, whereas the other gives correction induced by non-vanishing value of $\Phi$. 

\begin{figure*}
\centering
\includegraphics[width=0.8\textwidth]{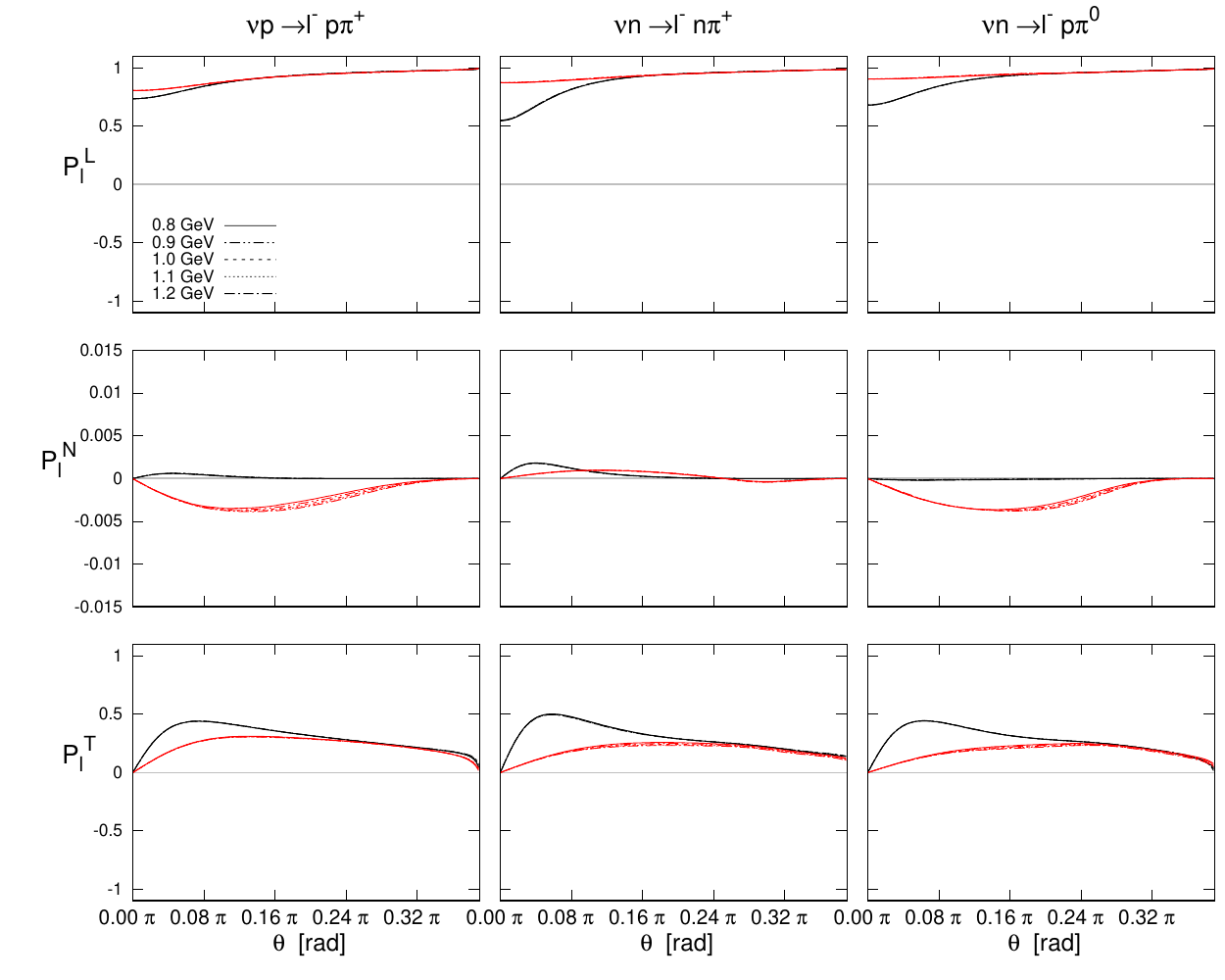}\\
\includegraphics[width=0.8\textwidth]{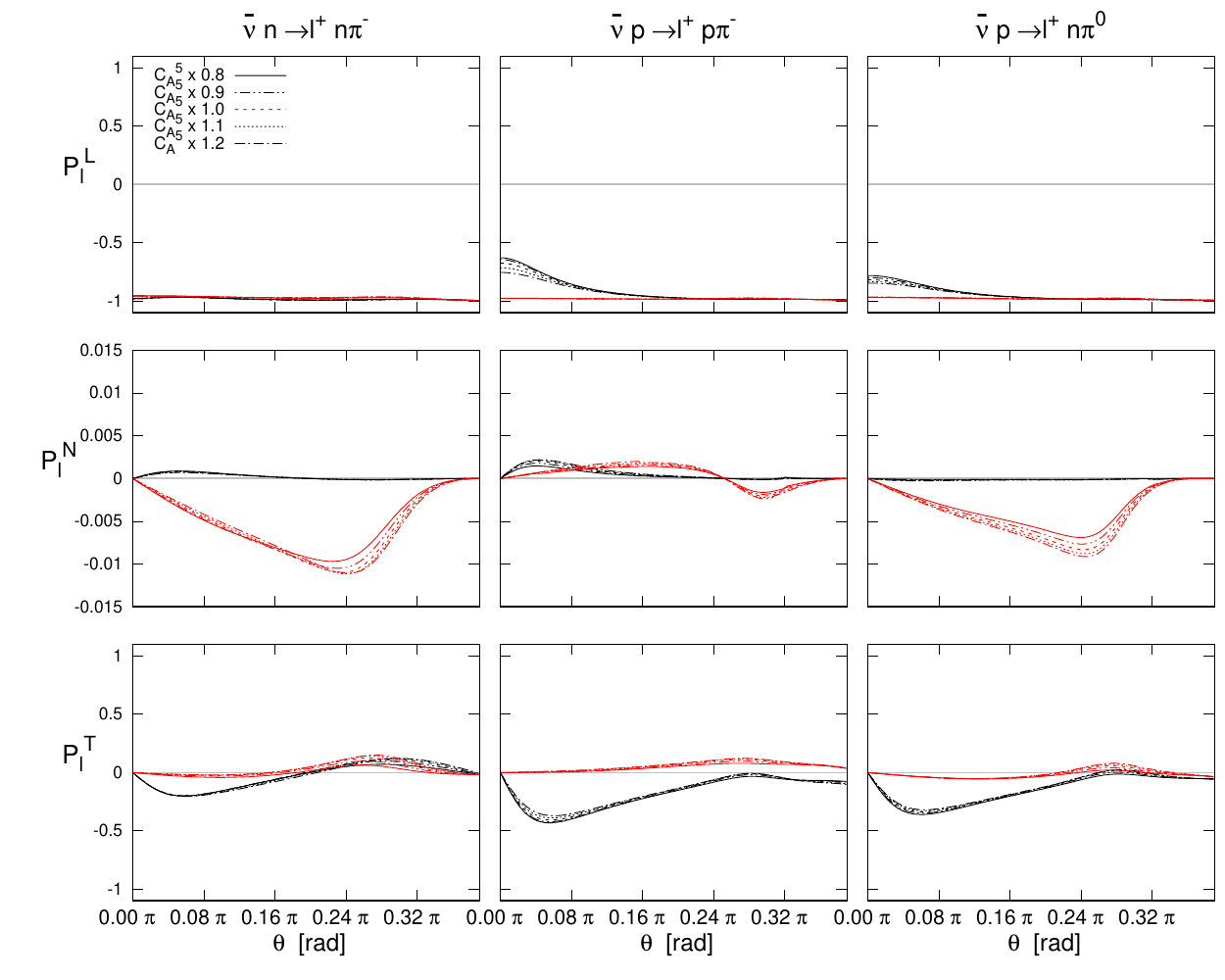}\\
\caption{The dependence of polarization of $\mu^-$ (top three rows) and $\mu^+$ (bottom three rows) on scattering angle $\theta$. Longitudinal $P_l^L$, normal  $P_l^N$ and transverse  $P_l^T$ components of the polarization of muon are calculated in the $HNV$ (black) and $FN$ (red) models, respectively for the energy $E=1$~GeV, and energy transfer $\omega=0.5$~GeV. In the top figures, $M_A$ is varied, while in the bottom figures, the plots for various values of $C_5^A$ are shown.
		}
		\label{fig:MA_lepton_polarization_E=1_w=0.5_neutrino}
\end{figure*}

\begin{figure*}
\centering
\includegraphics[width=0.8\textwidth]{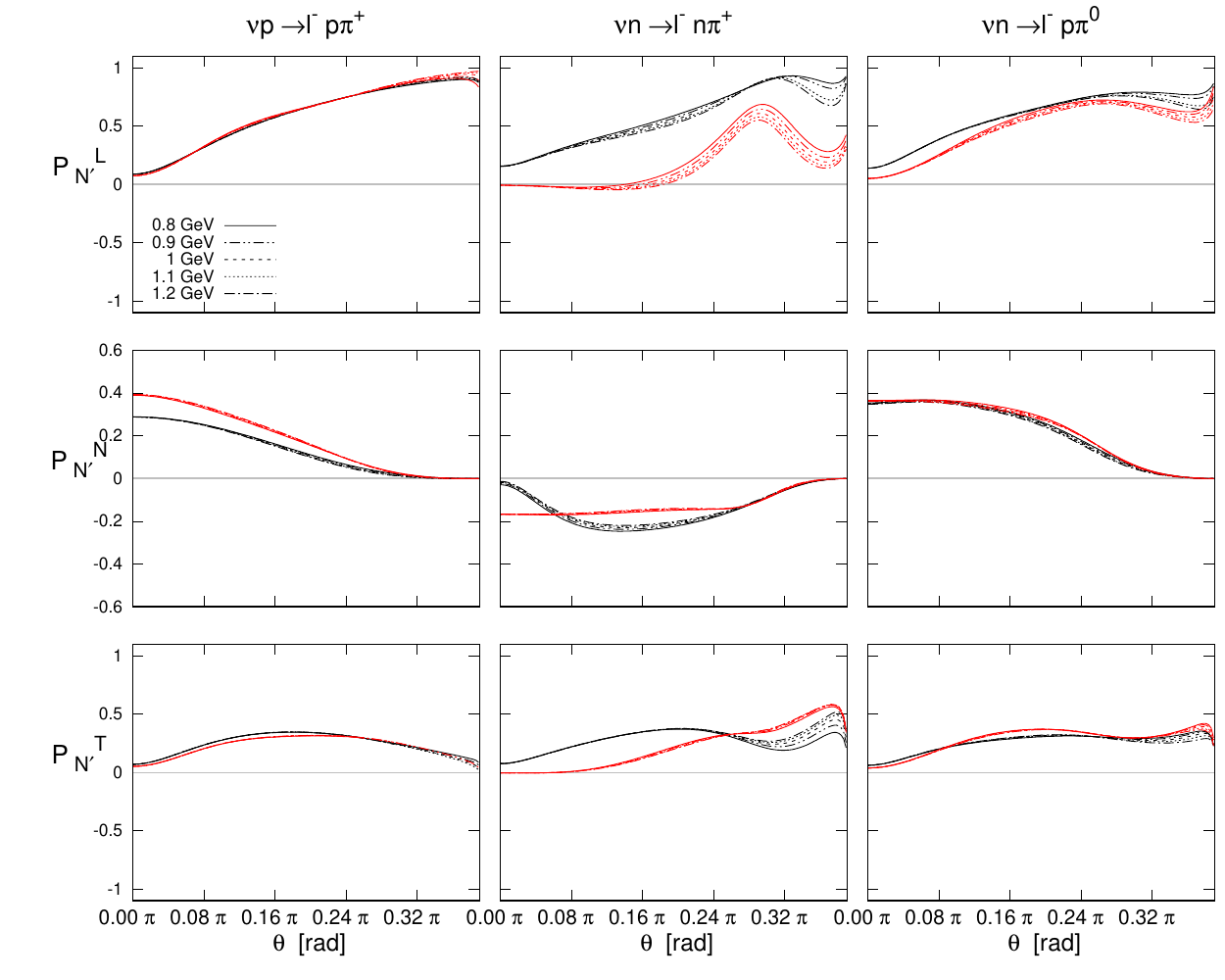}\\
\includegraphics[width=0.8\textwidth]{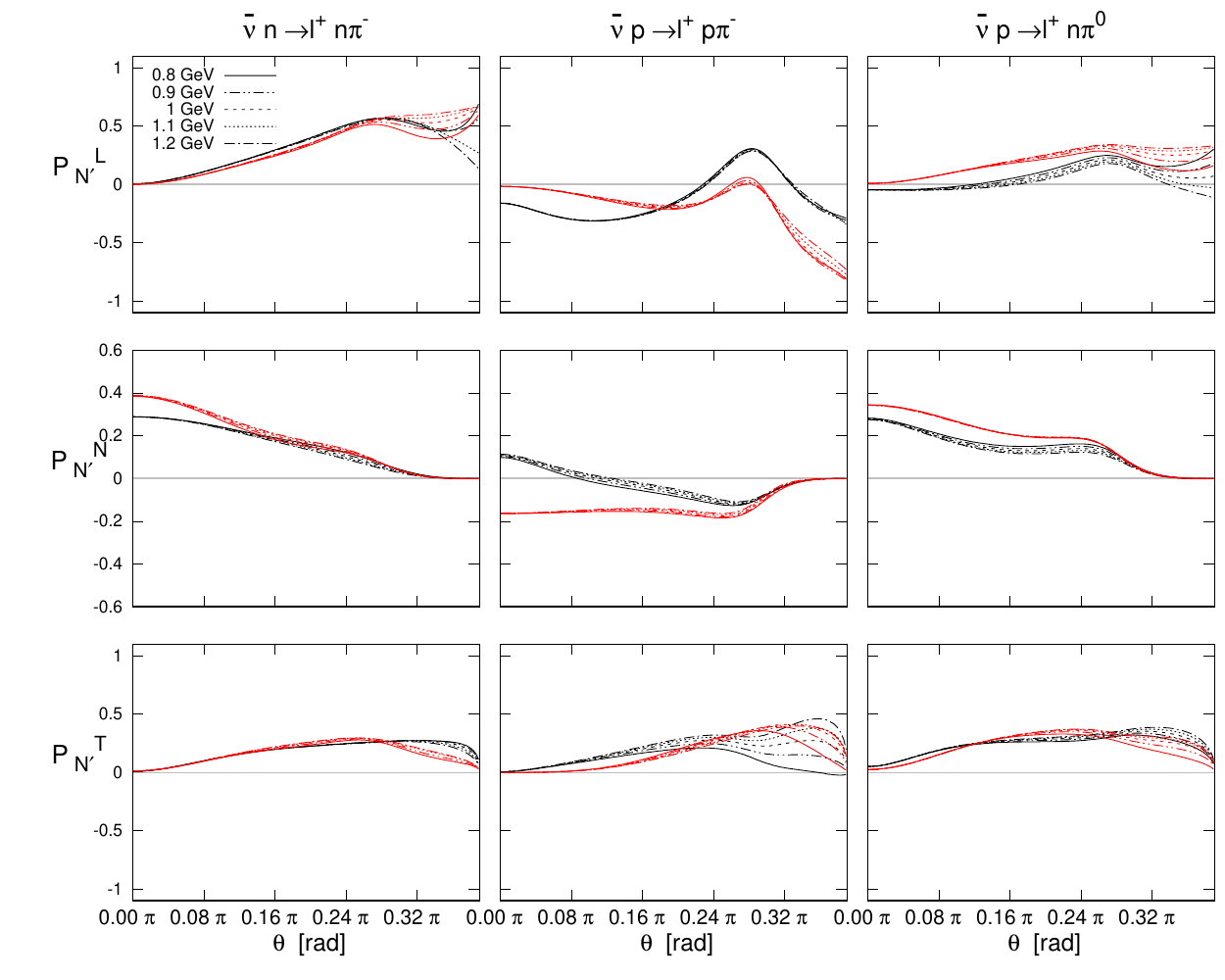}
\caption{
The dependence of polarization of final nucleon on scattering angle $\theta$. Longitudinal $P_{N'}^L$, normal  $P_{N'}^N$ and transverse  $P_{N'}^T$ components of the polarization are calculated in the $HNV$ (black) and $FN$ (red) models, respectively for the energy $E=1$~GeV, and energy transfer $\omega=0.5$~GeV. The axial mass is varied.
}
\label{fig:MA_nucleon_polarization_E=1_w=0.5_antyneutrino}
\end{figure*}
\begin{figure*}
\centering
\includegraphics[width=0.8\textwidth]{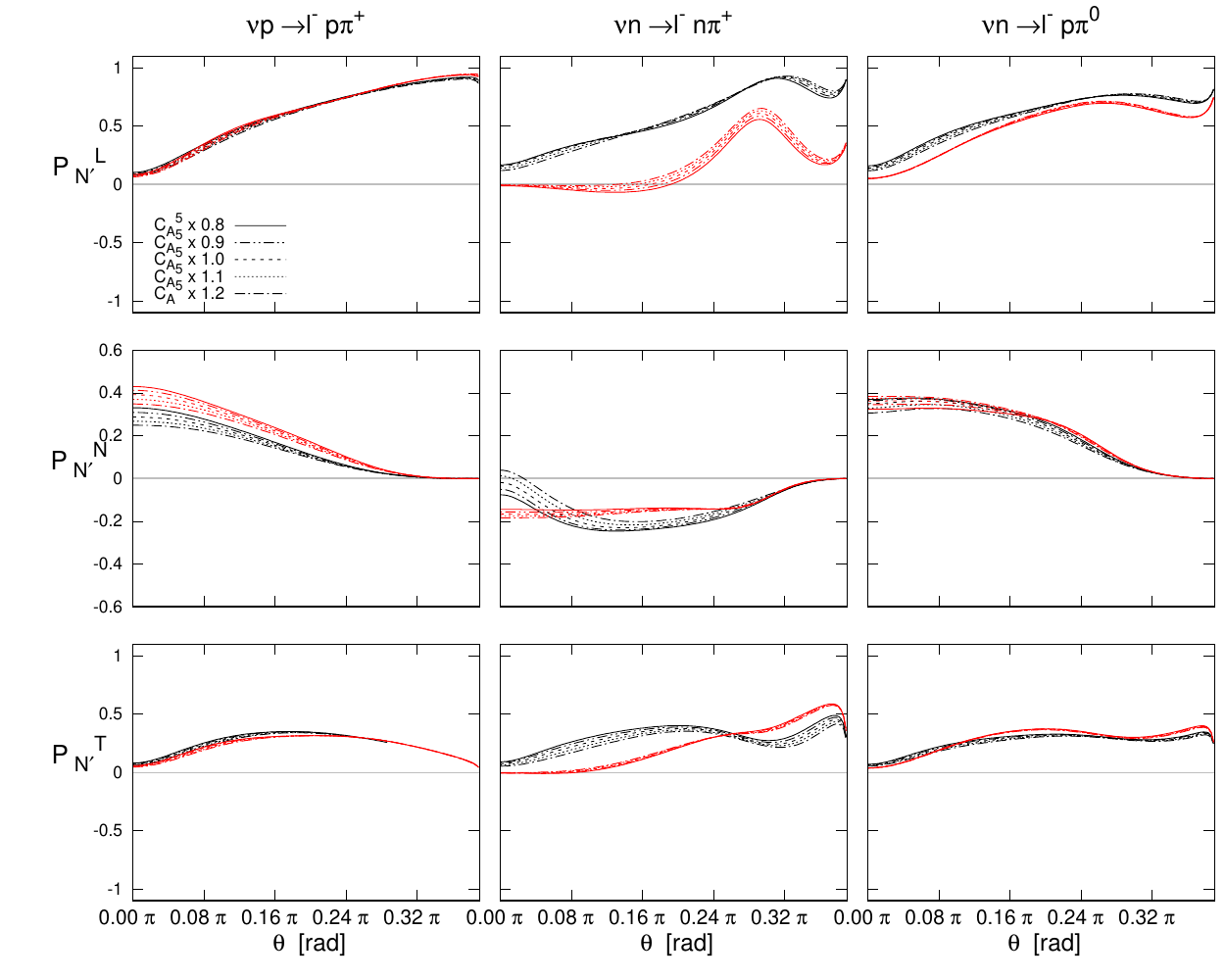}\\
\includegraphics[width=0.8\textwidth]{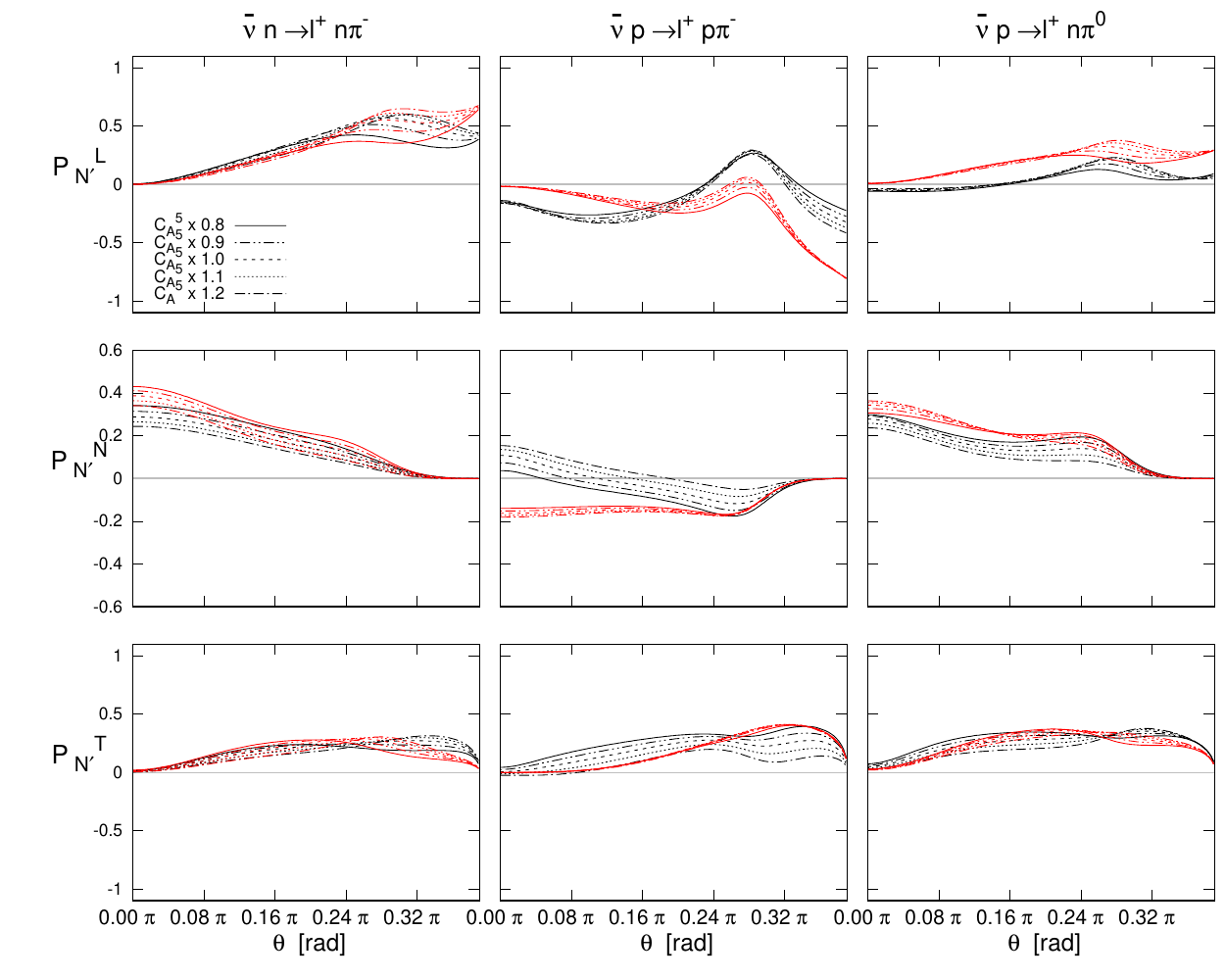}
\caption{
The dependence of polarization of final nucleon on scattering angle $\theta$. Longitudinal $P_{N'}^L$, normal  $P_{N'}^N$ and transverse  $P_{N'}^T$ components of the polarization are calculated in the $HNV$ (black) and $FN$ (red) models, respectively for the energy  $E=1$~GeV, and energy transfer $\omega=0.5$~GeV. The value of $C_5^A(0)$ is varied.
}
\label{fig:C5A_nucleon_polarization_E=1_w=0.5_antyneutrino}
\end{figure*}
\begin{figure*}
\centering 
\includegraphics[width=\textwidth]{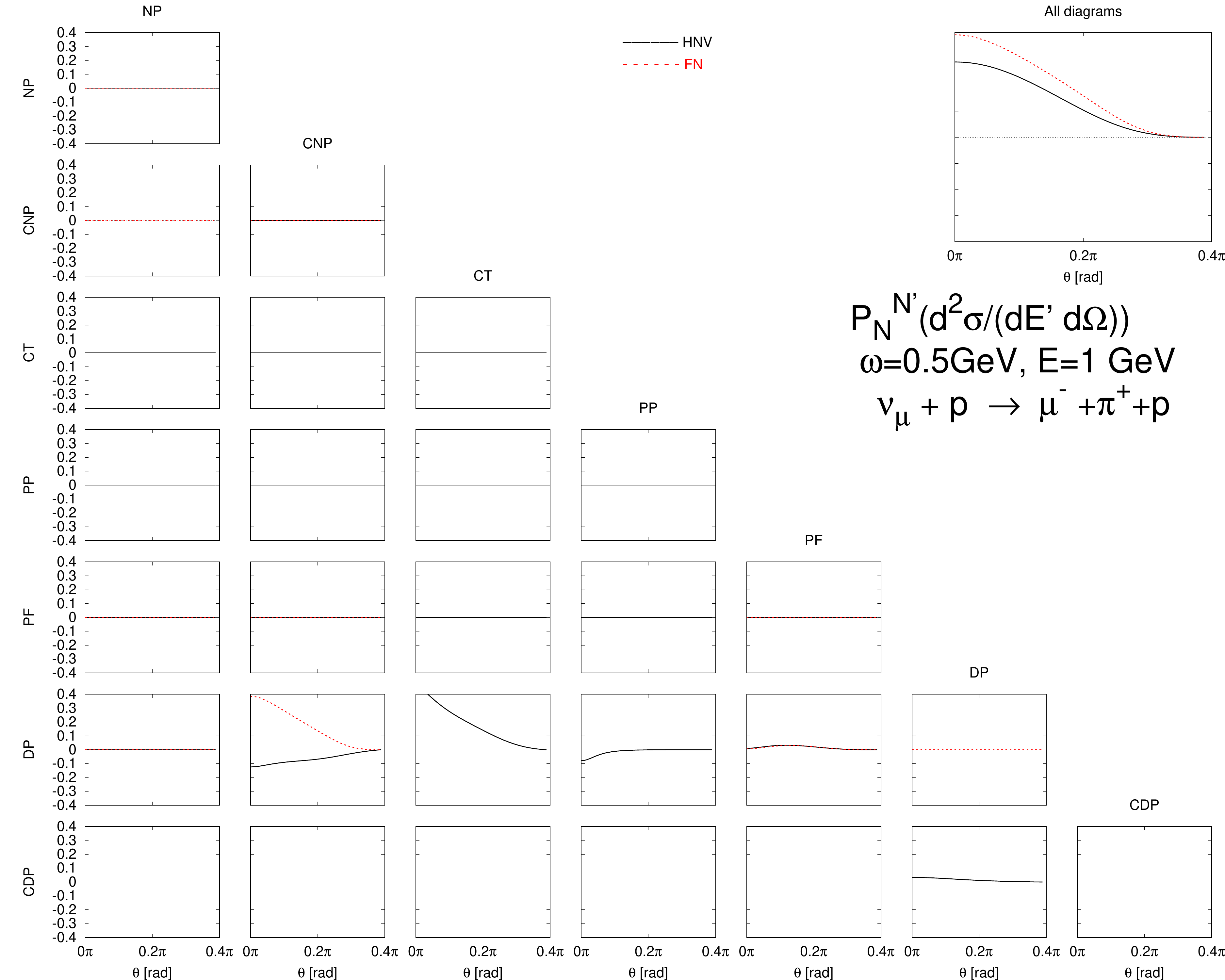}
\caption{
		 Separation of the normal component of the polarization of the recoil nucleon $P_{N'}^N$ into various interference contributions. Energy of the  neutrino is equal $E=1$~GeV, energy transfer is $\omega=0.5$~GeV. The phase between resonance and non-resonant background equal $0^\circ$. The dotted/solid(black/red) line represents the $HNV$/$FN$ model predictions. The abbreviations denote the  diagrams:  $DP$ \textit{-- delta pole} (diagram (f) in Fig.~\ref{fig:model-diagrams}), $CDP$ \textit{--  crossed delta pole} (diagram  (g)),  $NP$ \textit{-- nucleon pole} (diagram (a)), $CNP$ \textit{--  crossed nucleon pole} (diagram (b)), $CT$ \textit{-- contact term} (diagram (d)), $PP$ \textit{--  pion pole} (diagram (e)), $PF$ \textit{-- pion in flight (diagram (c))}. If $\mathcal{M}_a$ denotes the amplitude from given diagram, then  contribution from the $|\mathcal{M}_a|^2$ is on the diagonal, while below the diagonal the interference terms $2 \mathrm{Re}(\mathcal{M}_a\mathcal{M}_b)$ are plotted ($a$ indicates the column and $b$ the row).
		 \label{GFig:trojkatny}
	}
	\end{figure*}

\begin{figure*}
\centering
\includegraphics[width=0.8\textwidth]{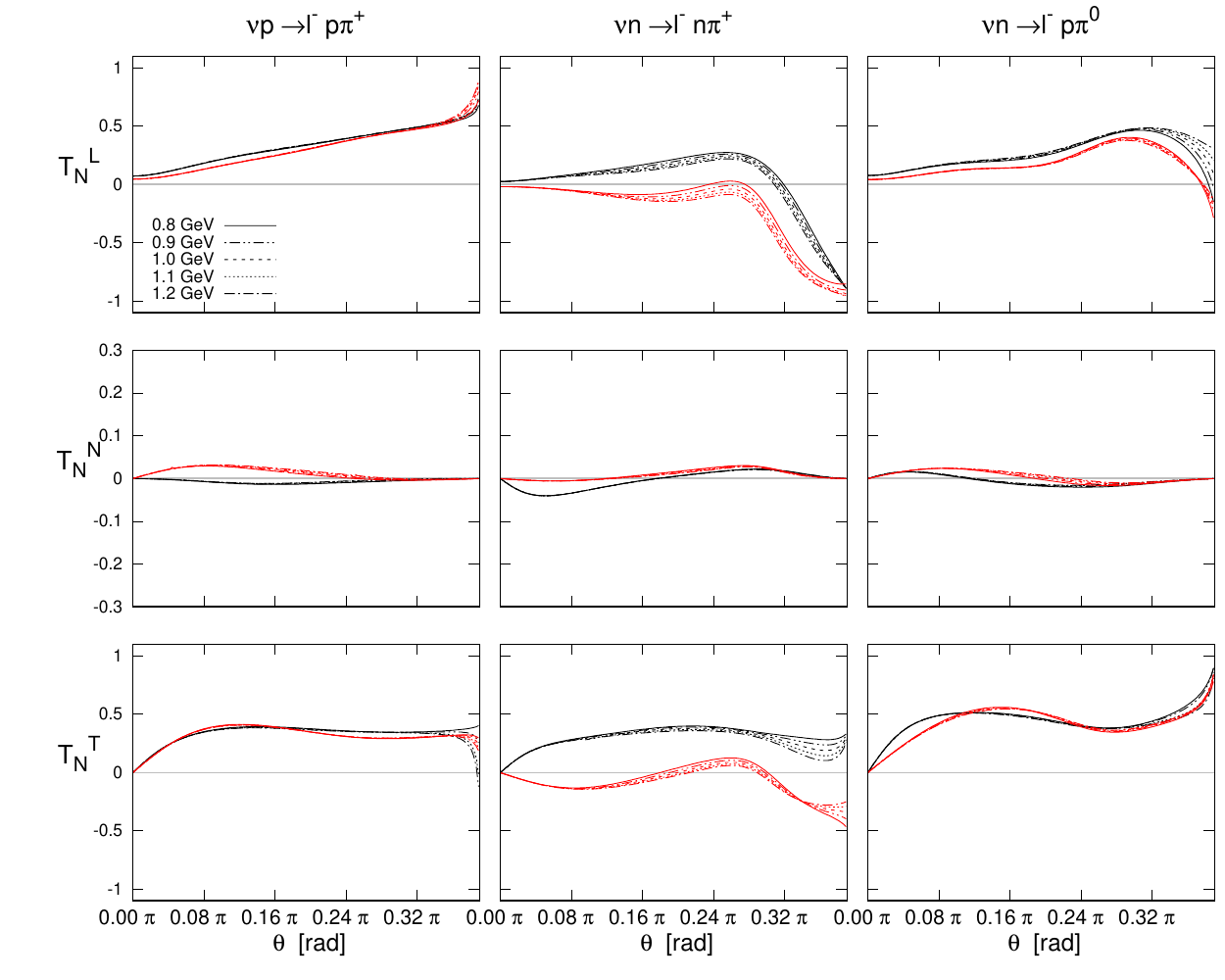}\\
\includegraphics[width=0.8\textwidth]{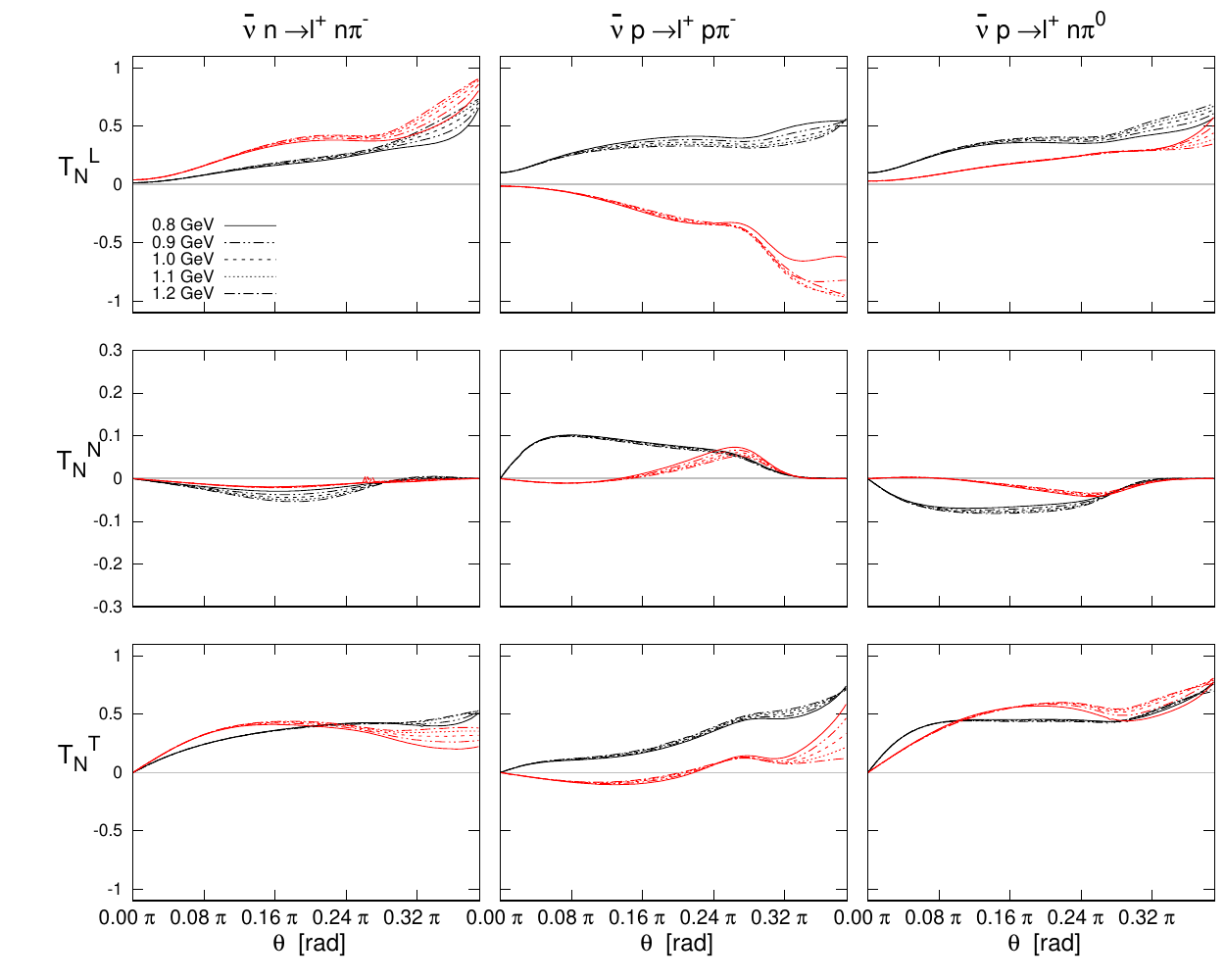}
\caption{
Dependence of the target asymmetry on scattering angle $\theta$ for $\nu_\mu N$ (tree top rows) and $\overline{\nu} N$ interactions (three bottom rows). Longitudinal $T_N^L$, normal  $T_N^N$ and transverse  $T_N^T$  components of the target asymmetry calculated in the $HNV$ (black) and $FN$ (red) model, respectively for the neutrino $E=1$~GeV, energy transfer $\omega=0.5$~GeV. The plots obtained for various values of $M_A$.
}
\label{fig:MA_target_polarization_E=1_w=0.5_neutrino}
\end{figure*}

\section{Numerical results and summary}
\label{sec:Results}

We consider six CC SPP processes, namely:
\begin{eqnarray}
 \label{Eq:processC0}   \nu_\mu + p        & \to & \mu^- + p + \pi^+ 
 \\
 \label{Eq:processC1} \nu_\mu + n          & \to & \mu^- + n + \pi^+
 \\
 \label{Eq:processC2}  \nu_\mu + n         & \to & \mu^- + p + \pi^0
 \\
\label{Eq:processC3}  \bar{\nu}_{\mu} + n  & \to & \mu^+ + n +\pi^-
\\
 \label{Eq:processC4}  \bar{\nu}_{\mu} + p & \to & \mu^+ + p +\pi^- 
\\
\label{Eq:processC5}  \bar{\nu}_{\mu} + p & \to & \mu^+ + n +\pi^0. 
\end{eqnarray}
We compute longitudinal, transverse, and normal polarization components of the final nucleon and the muon for each process. Similarly, three components of target asymmetry are presented.  In all numerical experiments, we limit the hadronic invariant mass to $W<1.4$~GeV.

In order to examine the sensitivity of the spin observables on the axial form factor $C_5^A$, we shall vary the parameters  $M_A$ and $C_5^A(0)$ in the dipole parametrization (\ref{Eq:C5A_dipole}).  The variation of $M_A$ mimic change of  shape of $C_5^A$, whereas increasing/decreasing $C_5^A(0)$ change the strength of $C_5^A$.  For numerical investigations we keep the following default values for the axial form factor parameters:
\begin{eqnarray}
\label{Eq:default_MA}
M_A  &=&   1.0~\mathrm{GeV}, \\
\label{Eq:default_C5A0_HNV}
C_5^A(0) &=& 1.10, \quad \mathrm{for}\,\mathrm{HNV}\,\mathrm{model}, \\
\label{Eq:default_C5A0_FN}
C_5^A(0) &=& 1.18, \quad \mathrm{for}\,\mathrm{FN}\,\mathrm{model}.
\end{eqnarray}
If one of the parameters is varied, the other keeps its default value.

Let us start the discussion from the analysis of the polarization of muon.
We immediately conclude that the muon polarization is not well suited for testing the hadronic model details. Indeed in Fig.~\ref{fig:MA_lepton_polarization_E=1_w=0.5_neutrino} (first three rows) we show the $M_A$-dependence of the polarization components of $\mu^-$, produced in $\nu_\mu N$. In the same figure (bottom three rows), we present the dependence of polarization of $\mu^+$, produced in $\overline{\nu}_\mu N$, on $C_5^A(0)$. We see that muon polarization is insensitive to axial form factor parametrization.   Some dependence on $M_A$ and $C_5^A(0)$ is visible for normal polarization of $\mu$. However, this component is small. It is no surprise because the muon is a light particle. Therefore it is almost completely longitudinally polarized.

 More interesting results are obtained for the final nucleon, see Figs.~\ref{fig:MA_nucleon_polarization_E=1_w=0.5_antyneutrino} and \ref{fig:C5A_nucleon_polarization_E=1_w=0.5_antyneutrino}. The polarization components show visible dependence on the $C_5^A(0)$ and $M_A$. But the polarization of the final nucleon is more sensitive to $C_5^A(0)$ than to $M_A$.  The effect is similar for both implemented SPP models.  
 Let us underline that we have chosen the kinematic configuration in which the normal component of the polarization takes large values. Indeed $\mathcal{P}_{N'}^N$ takes the value of about $0.4$ (for HNV) and $0.3$ (for FN) at a slight scattering angle. The normal component is determined entirely by the interference between $\mathcal{M}_R$ and $\mathcal{M}_{NB}$ amplitudes. This property is illustrated in  Fig.~\ref{GFig:trojkatny}, where we plot the normal component of the final nucleon and all contributing interference terms.

\begin{figure*}
\centering
\includegraphics[width=0.8\textwidth]{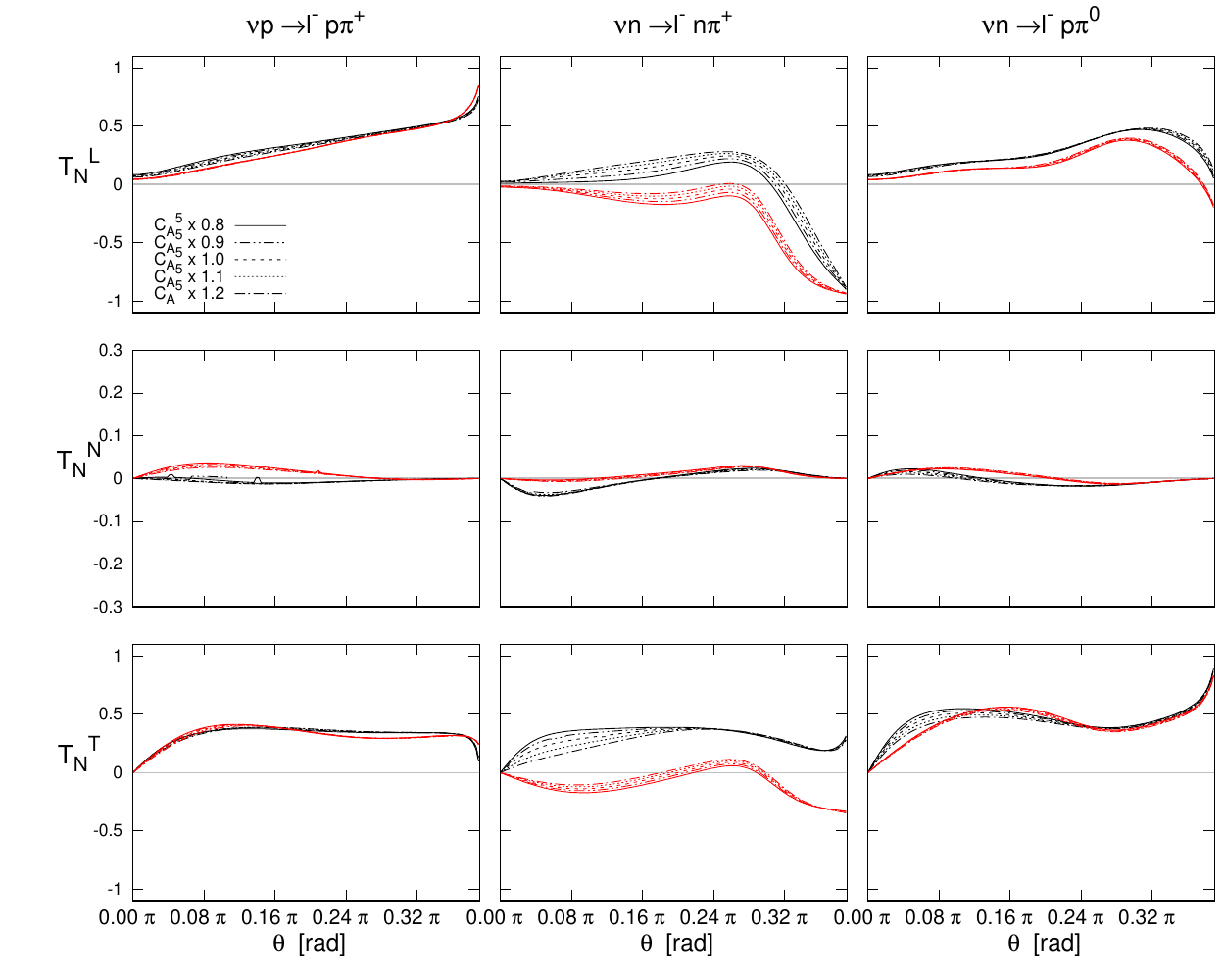}\\
\includegraphics[width=0.8\textwidth]{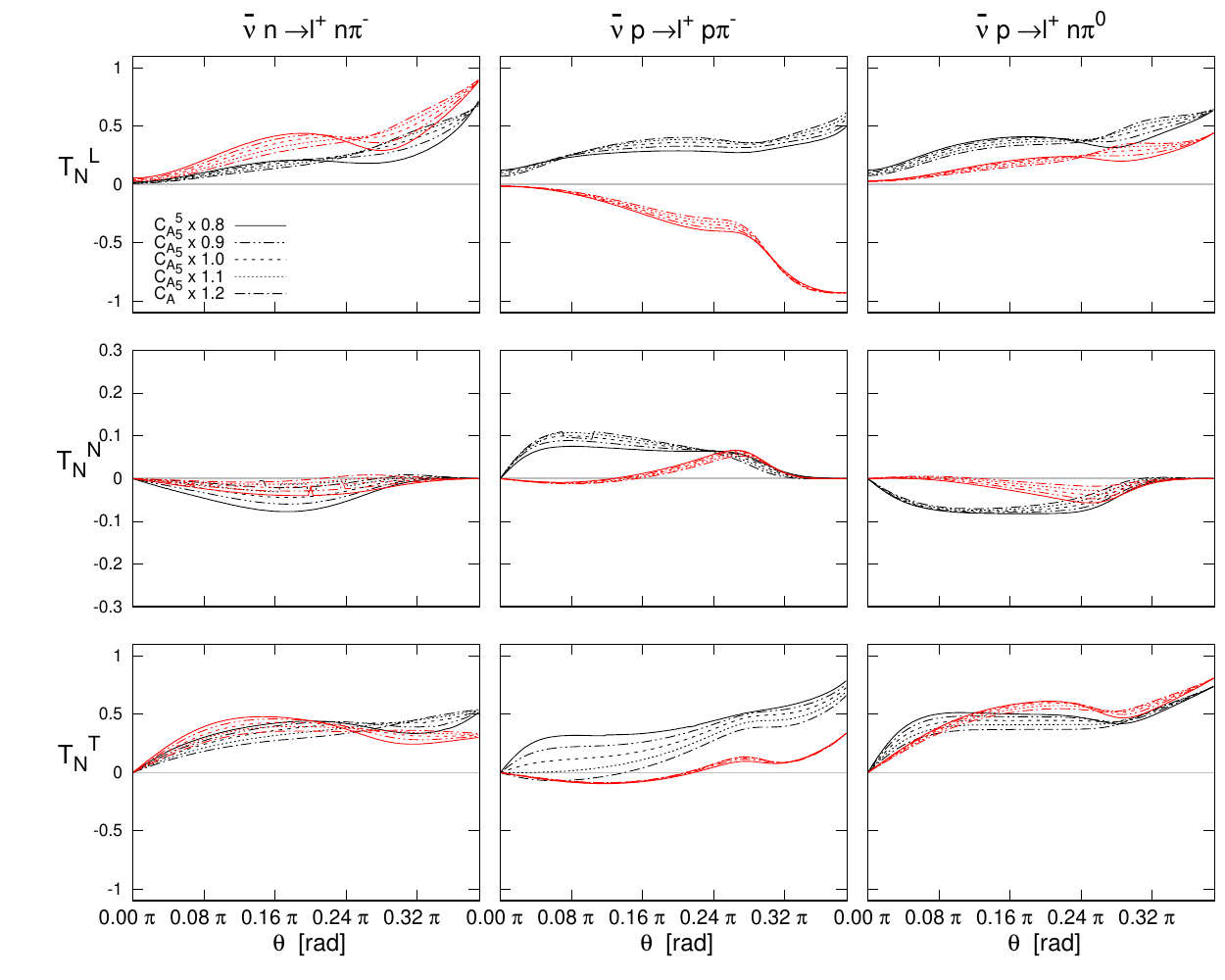}
\caption{
Dependence of the target asymmetry on scattering angle $\theta$ for $\nu_\mu N$ (tree top rows) and $\overline{\nu} N$ interactions (three bottom rows).
Longitudinal $T_N^L$, normal  $T_N^N$, and transverse  $T_N^T$ components of the target asymmetry calculated in the $HNV$ (black) and $FN$ (red) model, respectively for the neutrino $E=1$~GeV, energy transfer $\omega=0.5$~GeV. The plots are obtained for various values of  $C_5^A(0)$.
}
\label{fig:C5A_target_polarization_E=1_w=0.5_neutrino}
\end{figure*}

 The target asymmetry turns out to be the most sensitive to the variation of $C_5^A$, see Figs.~\ref{fig:MA_target_polarization_E=1_w=0.5_neutrino} and \ref{fig:C5A_target_polarization_E=1_w=0.5_neutrino}. The highest effect due to the change of the axial form factor parameters is seen for the longitudinal and the transverse components of target spin asymmetry computed for the antineutrino-nucleon scattering. 

Both discussed SPP models, as they were proposed initially, are non-unitary. However, the model HNV is more consistent in the formulation. Indeed, it consists of all required by gauge invariance diagrams (on tree level). Furthermore, the nonresonant background form factors are constrained by relation deduced from the vector current conservation.  Moreover, an effort has been made to unitarize the HNV model~\cite{Alvarez-Ruso:2015eva}. Therefore, in the rest of the paper, we shall only concentrate on the HNV model results. 

Below, we shall demonstrate that the polarization asymmetries are good observables for studying the unitarization procedures. Moreover, the measurement of them can allow to fixing the relative phases between amplitudes.  In order to perform the numerical experiments, we vary the phase $\Phi$ parameter. 

\begin{figure*}
		\centering
		\includegraphics[width=0.8\textwidth]{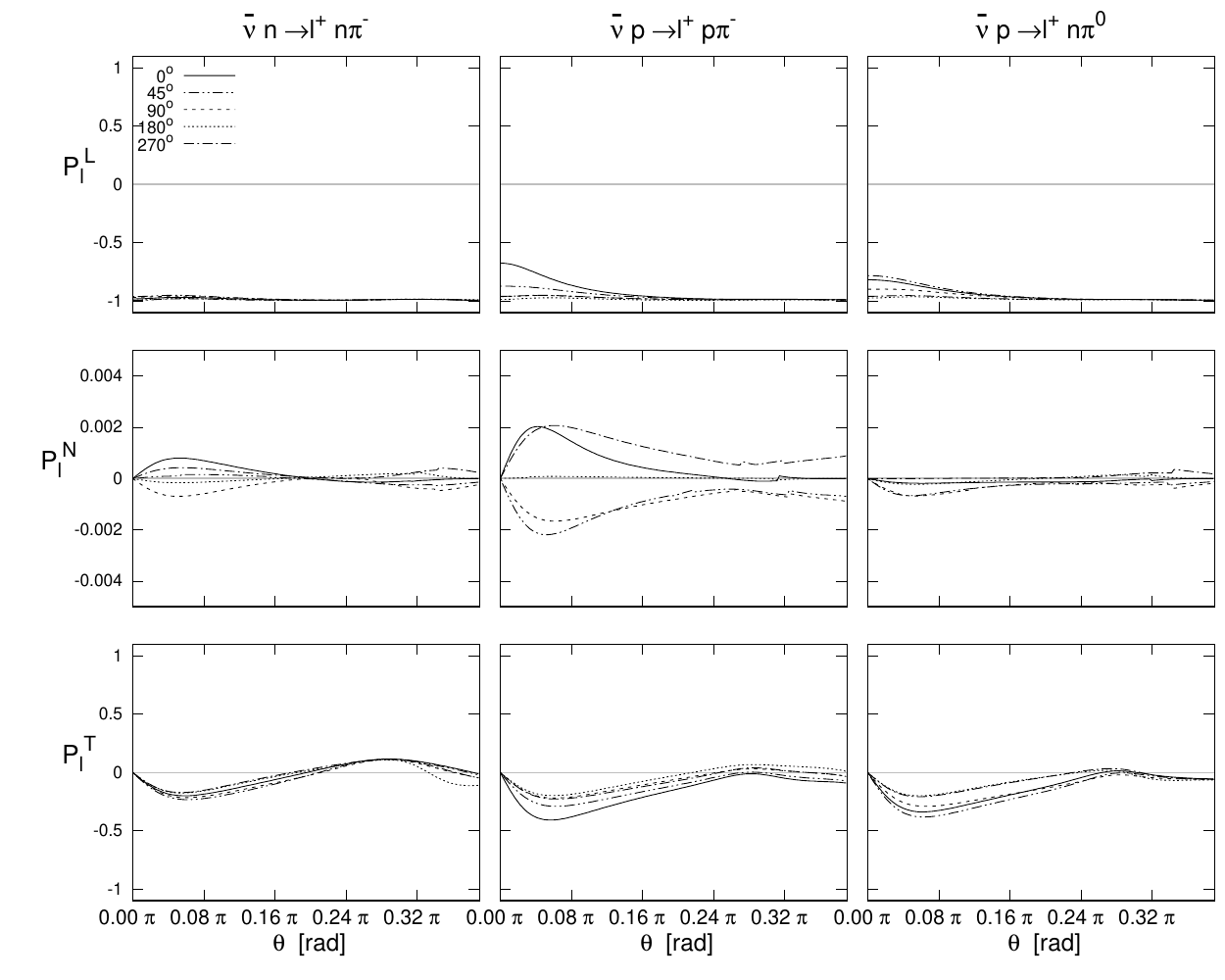} 
\caption{
The dependence of $\mu^+$ polarization on the scattering angle $\theta$. Polarization calculated within the $HNV$ model. Energy of the  neutrino $E=1$~GeV, energy transfer $\omega=0.5$~GeV. Solid, dotted-dotted-dashed, dashed, dotted, and dotted-dashed lines correspond to phases between resonance and non-resonant background equal $0^\circ,45^\circ,90^\circ, 180^\circ,270^\circ$, respectively.
		}
		\label{fig:PHASE_lepton_polarization_E=1_w=0.5_HNV}
\end{figure*}

Firstly, in Fig.~\ref{fig:PHASE_lepton_polarization_E=1_w=0.5_HNV} we present the dependence of the $\mu^+$ polarization on the phase factor. The longitudinal and transverse components of the muon polarization show some dependence on the phase factor. As expected \cite{Graczyk:2017rti}, the normal component depends strongly on  $\Phi$, but its absolute value is small.  
 
\begin{figure*}
		\centering
		\includegraphics[width=0.8\textwidth]{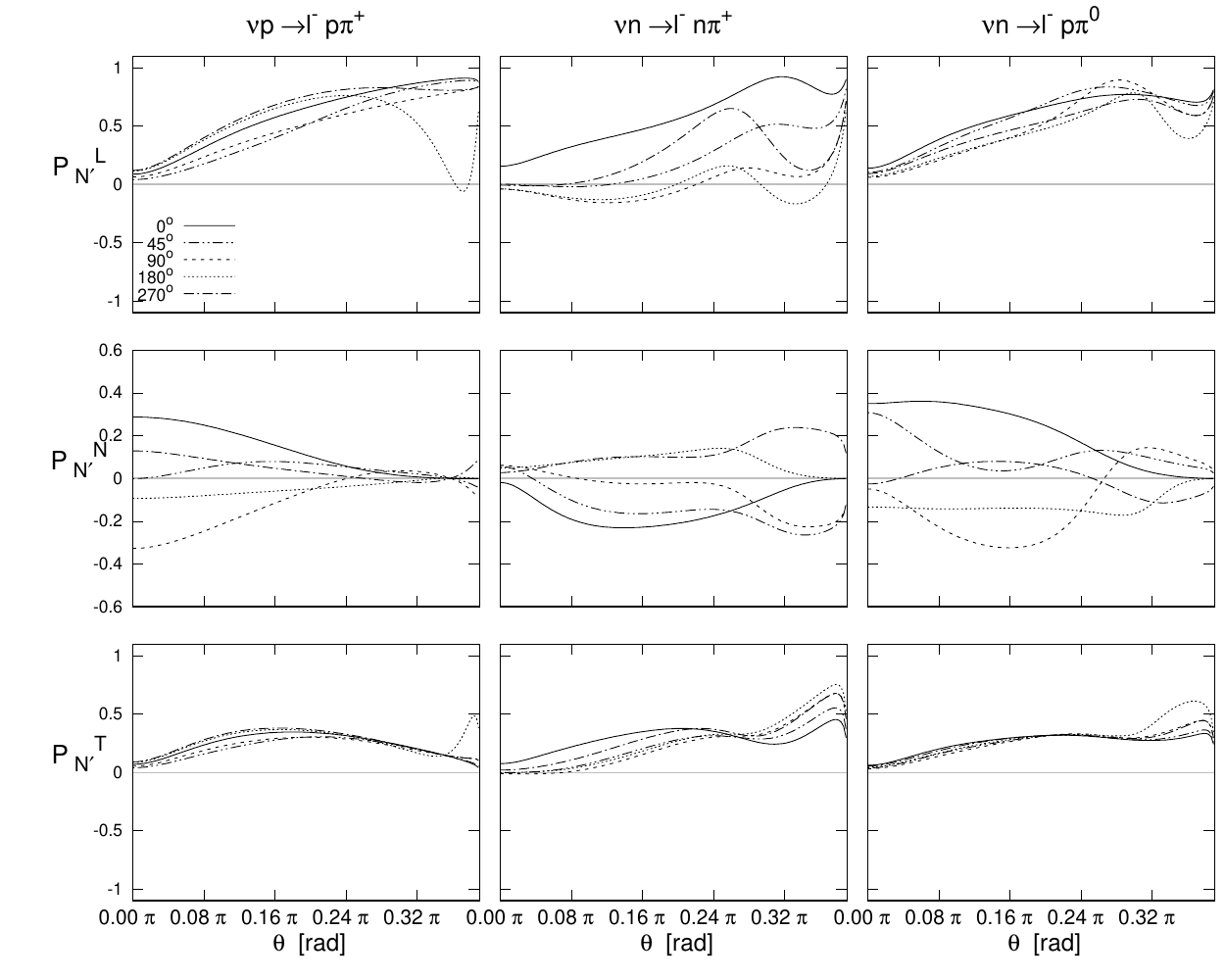}
		\includegraphics[width=0.8\textwidth]{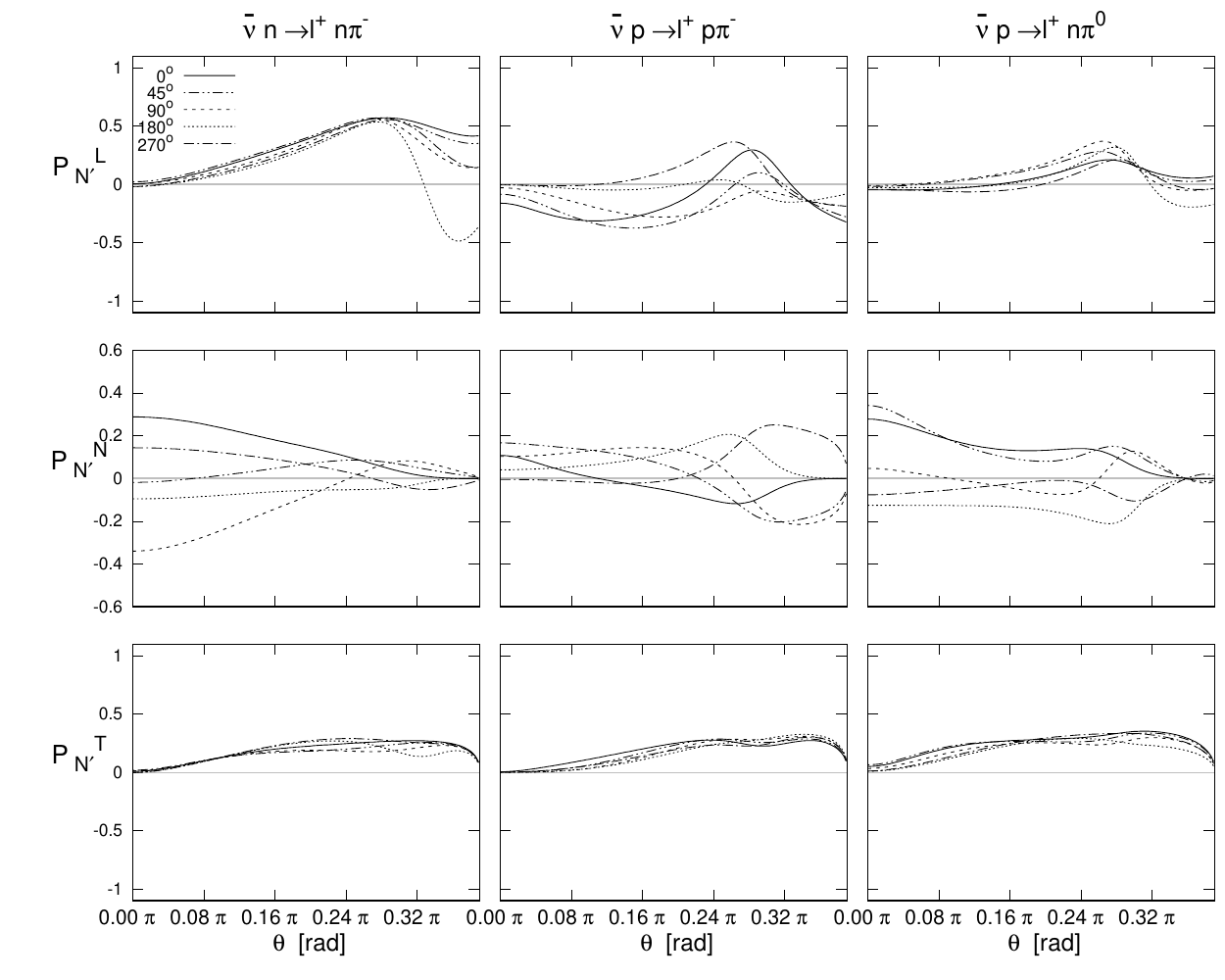}
		\caption{
	Dependence of the final nucleon polarization on the scattering angle $\theta$. Components $P_{N'}^L$, $P_{N'}^N$, and $P_{N'}^T$ of the final nucleon asymmetry calculated within the $HNV$ model. Energy of the  neutrino $E=1$~GeV, energy transfer $\omega=0.5$~GeV. Solid, dotted-dotted-dashed, dotted, dashed, and dotted-dashed lines correspond to phases between resonance and non-resonant background equal $0^\circ,45^\circ,90^\circ, 180^\circ,270^\circ$, respectively.
		}
	\label{fig:PHASE_recoil_polarization_E=1_w=0.5_HNV}
	\end{figure*}

\begin{figure*}
		\centering
		\includegraphics[width=0.8\textwidth]{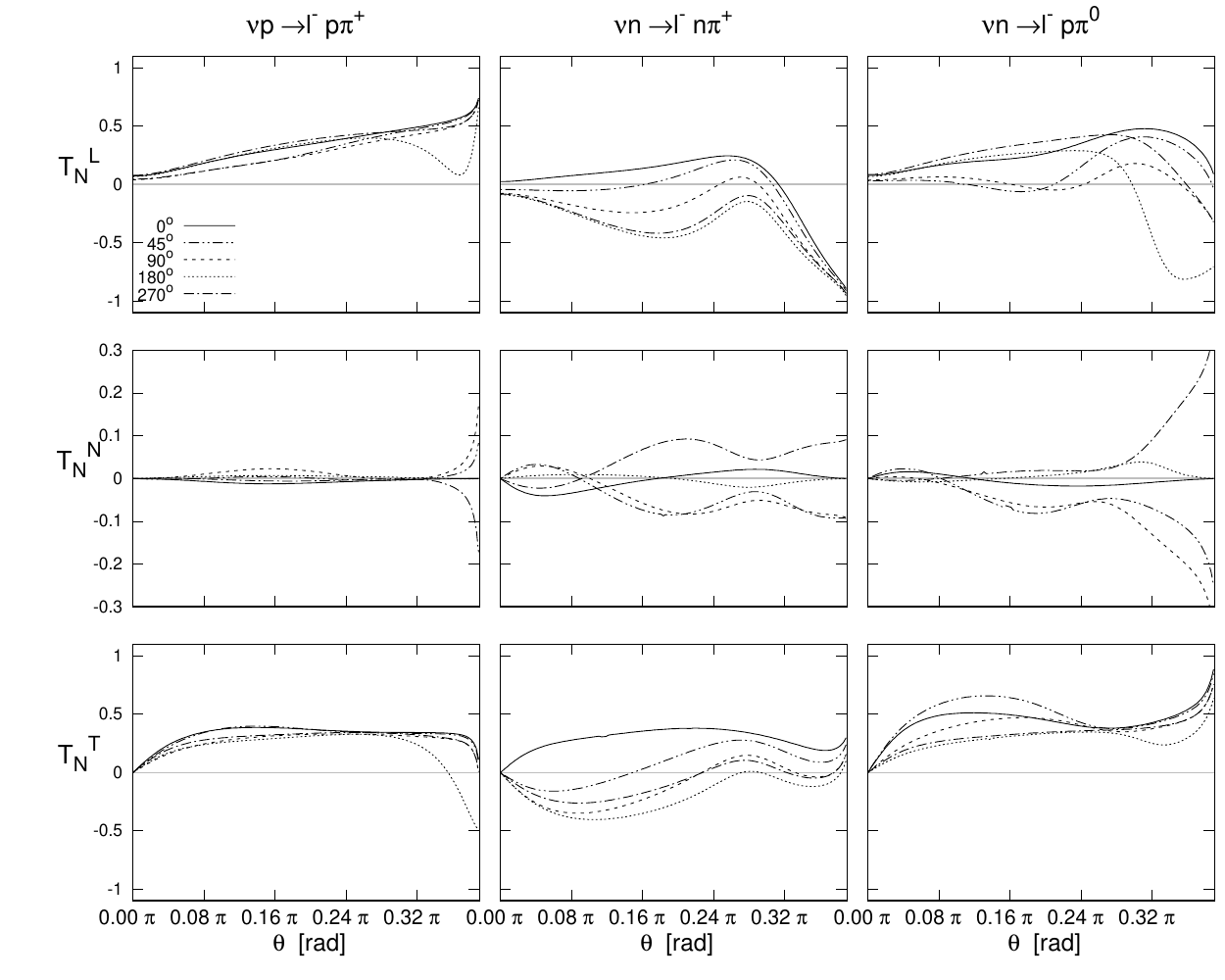}
		\includegraphics[width=0.8\textwidth]{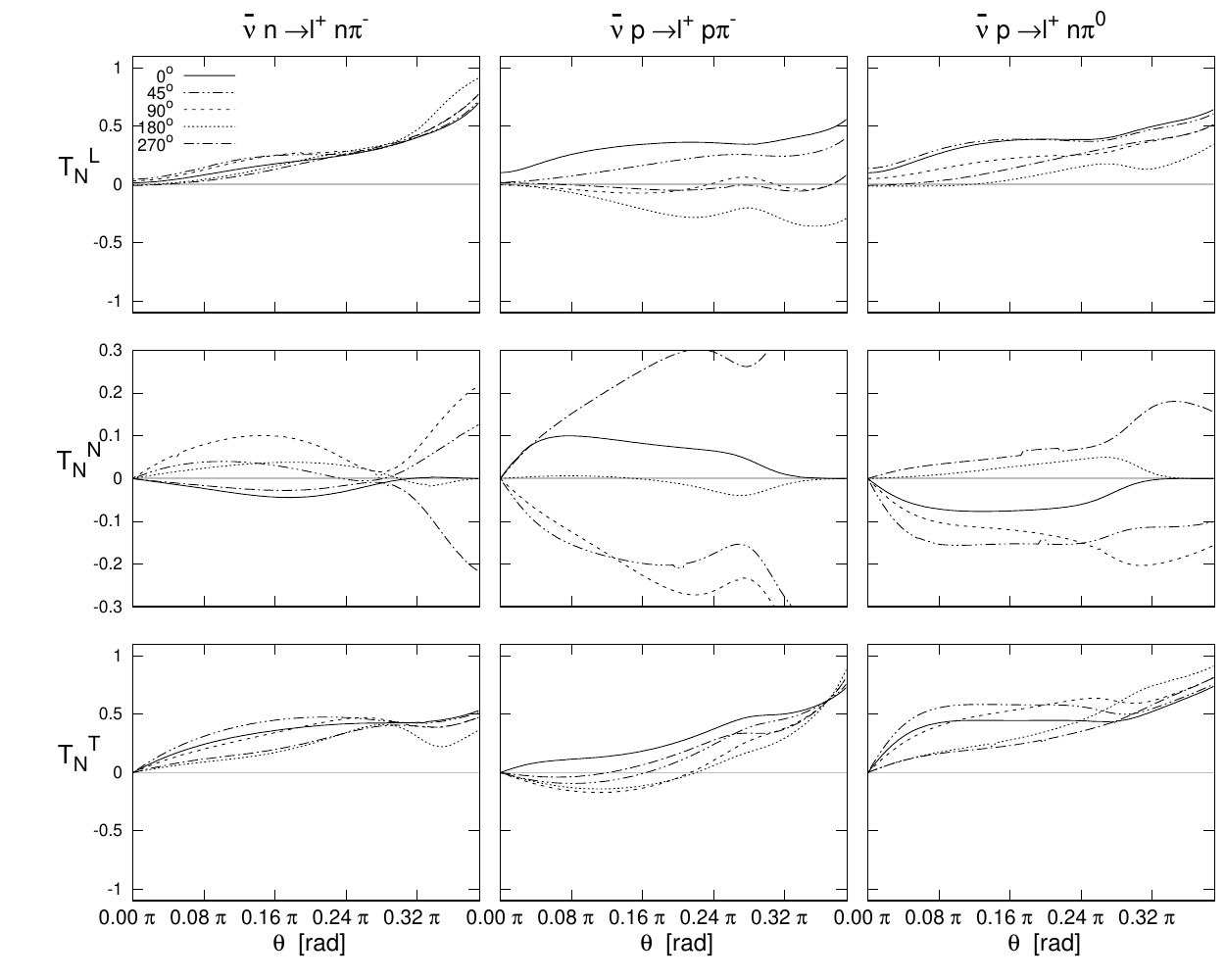}
		\caption{
				Dependence of the target spin asymmetry on the scattering angle $\theta$.
			Components $T_N^L$, $T_N^N$, and $T_N^T$ of the target asymmetry calculated within the $HNV$ model. Energy of the  neutrino $E=1$~GeV, energy transfer $\omega=0.5$~GeV. Solid, dotted-dotted-dashed, dashed, dotted, and dotted-dashed lines correspond to phases between resonance and non-resonant background equal $0^\circ,45^\circ,90^\circ, 180^\circ,270^\circ$, respectively.
		}
	\label{fig:PHASE_target_polarization_E=1_w=0.5_HNV}
	\end{figure*}

Secondly, in Figs.~\ref{fig:PHASE_recoil_polarization_E=1_w=0.5_HNV} and \ref{fig:PHASE_target_polarization_E=1_w=0.5_HNV} we present the $\Phi$-dependence of the final nucleon polarization and the target spin asymmetry. It can be noticed that the final nucleon's polarization and the target spin asymmetry are well suited to test the unitarization procedures.  Measurement of them should help in establishing the phase factors. The longitudinal and normal components of the final nucleon and all components of the target spin asymmetry are the most sensitive to the change of the phase factor. Indeed, the phase impacts the sign and the absolute value of these components.


Summary: in this work, we have extended our previous investigations of properties of the spin asymmetries, in CC SPP neutrino nucleon scattering, by discussing: 
\begin{itemize}
	\item the sensitivity of spin observables to the axial structure of the $\Delta(1232)$ resonance;
\item dependence of the spin asymmetries on the relative phase between resonance and nonresonance amplitudes.
\end{itemize}
We conclude that the final nucleon polarization and the target spin asymmetry are sensitive to the change of axial mass and $C_5^A(0)$ parameters. Hence the polarization observables and spin-averaged cross-section data should give complete information about the axial structure of $\Delta(1232)$ resonance.  Moreover, both types of polarization observables are well suited for studies the unitarization procedures and fixing the relative phases between resonance-nonresonant amplitudes.

\vspace{1cm}

\begin{acknowledgments}

The calculations have been carried out in Wroclaw Centre for Networking and Supercomputing (\url{http://www.wcss.wroc.pl}), grant No. 268.\\
 
Research project partly supported by program ''Excellence initiative - research university'' for years 2020-2026 for University of Wroclaw.\\

A part of the algebraic calculations presented has been performed using FORM language \cite{Vermaseren:2000nd} and FeynCalc package~\cite{Mertig:1990an,Shtabovenko:2016sxi}.
 
\end{acknowledgments}

\appendix

\section{Spin Vector Basis}
\label{Appendix:Spin_baisis}

	\begin{itemize}
		\item Basis spin vectors for final lepton:
		\begin{eqnarray}
		\label{basis_lepton_L}
		\zeta_L^\mu    &=& \frac{1}{m } \left(|\textbf{k}'| ,  \frac{ E_{k'}\textbf{k}'}{ |\textbf{k}'|} \right) 
		\\
		\label{basis_lepton_T}
		\zeta_T^\mu        &=&  \left(0,\frac{\textbf{k}'\times(\textbf{k}\times \textbf{q})}{|\textbf{k}'\times(\textbf{k}\times \textbf{q})|}\right) 
		\\
		\label{basis_lepton_N}
		\zeta_N^\mu        &=& \left(0,\frac{\textbf{k}\times \textbf{q}}{|\textbf{k}\times \textbf{q}|} \right),
		\end{eqnarray}
		\item Basis spin vectors for final nucleon:
		\begin{eqnarray}
		\label{basis_recoil_L}
		\xi_L^\mu       & = &
		\frac{1}{M} \left(|\textbf{p}'|,\frac{ E_{p'} \textbf{p}' }{|\textbf{p}'|}\right),
		\\
		\xi_T^\mu       & = &  \left(0,\frac{\textbf{k}_{\pi}\times(\textbf{k}_{\pi}\times \textbf{q})}{|\textbf{k}_{\pi}\times(\textbf{k}_{\pi}\times \textbf{q})|}\right), 
		\\
		\xi_N^\mu       & = & 
		\left(0,\frac{\textbf{k}_{\pi}\times \textbf{q}}{|\textbf{k}_{\pi}\times \textbf{q}|} \right); 
		\end{eqnarray}
		\item Basis spin vectors for target:
		\begin{eqnarray} 
		\label{basis_target_L}
		\chi_L^\mu    &=& \frac{1}{E} \left(0,\textbf{k} \right), 
		\\
		\label{basis_target_T}
		\chi_T^\mu        &=& \left(0,\frac{\textbf{k}\times(\textbf{k}\times \textbf{q})}{|\textbf{k}\times(\textbf{k}\times \textbf{q})|}\right), \\
		\label{basis_target_N}
		\chi_N^\mu    &=&\left(0,\frac{\textbf{k}\times \textbf{q}}{|\textbf{k}\times \textbf{q}|}  \right).
		\end{eqnarray}
	\end{itemize}

\normalem
\bibliographystyle{apsrev4-1}
\bibliography{bibdrat,bibdratbook,bibmoje}

\begin{thebibliography}{108}%
\makeatletter
\providecommand \@ifxundefined [1]{%
 \@ifx{#1\undefined}
}%
\providecommand \@ifnum [1]{%
 \ifnum #1\expandafter \@firstoftwo
 \else \expandafter \@secondoftwo
 \fi
}%
\providecommand \@ifx [1]{%
 \ifx #1\expandafter \@firstoftwo
 \else \expandafter \@secondoftwo
 \fi
}%
\providecommand \natexlab [1]{#1}%
\providecommand \enquote  [1]{``#1''}%
\providecommand \bibnamefont  [1]{#1}%
\providecommand \bibfnamefont [1]{#1}%
\providecommand \citenamefont [1]{#1}%
\providecommand \href@noop [0]{\@secondoftwo}%
\providecommand \href [0]{\begingroup \@sanitize@url \@href}%
\providecommand \@href[1]{\@@startlink{#1}\@@href}%
\providecommand \@@href[1]{\endgroup#1\@@endlink}%
\providecommand \@sanitize@url [0]{\catcode `\\12\catcode `\$12\catcode
  `\&12\catcode `\#12\catcode `\^12\catcode `\_12\catcode `\%12\relax}%
\providecommand \@@startlink[1]{}%
\providecommand \@@endlink[0]{}%
\providecommand \url  [0]{\begingroup\@sanitize@url \@url }%
\providecommand \@url [1]{\endgroup\@href {#1}{\urlprefix }}%
\providecommand \urlprefix  [0]{URL }%
\providecommand \Eprint [0]{\href }%
\providecommand \doibase [0]{http://dx.doi.org/}%
\providecommand \selectlanguage [0]{\@gobble}%
\providecommand \bibinfo  [0]{\@secondoftwo}%
\providecommand \bibfield  [0]{\@secondoftwo}%
\providecommand \translation [1]{[#1]}%
\providecommand \BibitemOpen [0]{}%
\providecommand \bibitemStop [0]{}%
\providecommand \bibitemNoStop [0]{.\EOS\space}%
\providecommand \EOS [0]{\spacefactor3000\relax}%
\providecommand \BibitemShut  [1]{\csname bibitem#1\endcsname}%
\let\auto@bib@innerbib\@empty
\bibitem [{\citenamefont {Hernandez}\ \emph
  {et~al.}(2007{\natexlab{a}})\citenamefont {Hernandez}, \citenamefont
  {Nieves},\ and\ \citenamefont {Valverde}}]{Hernandez:2007qq}%
  \BibitemOpen
  \bibfield  {author} {\bibinfo {author} {\bibfnamefont {E.}~\bibnamefont
  {Hernandez}}, \bibinfo {author} {\bibfnamefont {J.}~\bibnamefont {Nieves}}, \
  and\ \bibinfo {author} {\bibfnamefont {M.}~\bibnamefont {Valverde}},\ }\href
  {\doibase 10.1103/PhysRevD.76.033005} {\bibfield  {journal} {\bibinfo
  {journal} {Phys. Rev.}\ }\textbf {\bibinfo {volume} {D76}},\ \bibinfo {pages}
  {033005} (\bibinfo {year} {2007}{\natexlab{a}})},\ \Eprint
  {http://arxiv.org/abs/hep-ph/0701149} {arXiv:hep-ph/0701149 [hep-ph]}
  \BibitemShut {NoStop}%
\bibitem [{\citenamefont {Fogli}\ and\ \citenamefont
  {Nardulli}(1979)}]{Fogli:1979cz}%
  \BibitemOpen
  \bibfield  {author} {\bibinfo {author} {\bibfnamefont {G.~L.}\ \bibnamefont
  {Fogli}}\ and\ \bibinfo {author} {\bibfnamefont {G.}~\bibnamefont
  {Nardulli}},\ }\href {\doibase 10.1016/0550-3213(79)90233-5} {\bibfield
  {journal} {\bibinfo  {journal} {Nucl. Phys.}\ }\textbf {\bibinfo {volume}
  {B160}},\ \bibinfo {pages} {116} (\bibinfo {year} {1979})}\BibitemShut
  {NoStop}%
\bibitem [{\citenamefont {Alvarez-Ruso}\ \emph {et~al.}(2014)\citenamefont
  {Alvarez-Ruso}, \citenamefont {Hayato},\ and\ \citenamefont
  {Nieves}}]{Alvarez-Ruso:2014bla}%
  \BibitemOpen
  \bibfield  {author} {\bibinfo {author} {\bibfnamefont {L.}~\bibnamefont
  {Alvarez-Ruso}}, \bibinfo {author} {\bibfnamefont {Y.}~\bibnamefont
  {Hayato}}, \ and\ \bibinfo {author} {\bibfnamefont {J.}~\bibnamefont
  {Nieves}},\ }\href {\doibase 10.1088/1367-2630/16/7/075015} {\bibfield
  {journal} {\bibinfo  {journal} {New J. Phys.}\ }\textbf {\bibinfo {volume}
  {16}},\ \bibinfo {pages} {075015} (\bibinfo {year} {2014})},\ \Eprint
  {http://arxiv.org/abs/1403.2673} {arXiv:1403.2673 [hep-ph]} \BibitemShut
  {NoStop}%
\bibitem [{\citenamefont {Mosel}(2016)}]{Mosel_annurev-nucl-102115-044720}%
  \BibitemOpen
  \bibfield  {author} {\bibinfo {author} {\bibfnamefont {U.}~\bibnamefont
  {Mosel}},\ }\href {\doibase 10.1146/annurev-nucl-102115-044720} {\bibfield
  {journal} {\bibinfo  {journal} {{Annual Review of Nuclear and Particle
  Science}}\ }\textbf {\bibinfo {volume} {66}},\ \bibinfo {pages} {171}
  (\bibinfo {year} {2016})}\BibitemShut {NoStop}%
\bibitem [{\citenamefont {Katori}\ and\ \citenamefont
  {Martini}(2018)}]{Katori:2016yel}%
  \BibitemOpen
  \bibfield  {author} {\bibinfo {author} {\bibfnamefont {T.}~\bibnamefont
  {Katori}}\ and\ \bibinfo {author} {\bibfnamefont {M.}~\bibnamefont
  {Martini}},\ }\href {\doibase 10.1088/1361-6471/aa8bf7} {\bibfield  {journal}
  {\bibinfo  {journal} {J. Phys. G}\ }\textbf {\bibinfo {volume} {45}},\
  \bibinfo {pages} {013001} (\bibinfo {year} {2018})},\ \Eprint
  {http://arxiv.org/abs/1611.07770} {arXiv:1611.07770 [hep-ph]} \BibitemShut
  {NoStop}%
\bibitem [{\citenamefont {Athar}\ and\ \citenamefont
  {Singh}(2020)}]{Athar:1970esm}%
  \BibitemOpen
  \bibfield  {author} {\bibinfo {author} {\bibfnamefont {M.~S.}\ \bibnamefont
  {Athar}}\ and\ \bibinfo {author} {\bibfnamefont {S.~K.}\ \bibnamefont
  {Singh}},\ }\href {\doibase 10.1017/9781108489065} {\emph {\bibinfo {title}
  {{The Physics of Neutrino Interactions}}}}\ (\bibinfo  {publisher} {Cambridge
  University Press},\ \bibinfo {year} {2020})\BibitemShut {NoStop}%
\bibitem [{\citenamefont {Sajjad~Athar}\ and\ \citenamefont
  {Morf\'\i{}n}(2021)}]{SajjadAthar:2020nvy}%
  \BibitemOpen
  \bibfield  {author} {\bibinfo {author} {\bibfnamefont {M.}~\bibnamefont
  {Sajjad~Athar}}\ and\ \bibinfo {author} {\bibfnamefont {J.~G.}\ \bibnamefont
  {Morf\'\i{}n}},\ }\href {\doibase 10.1088/1361-6471/abbb11} {\bibfield
  {journal} {\bibinfo  {journal} {J. Phys. G}\ }\textbf {\bibinfo {volume}
  {48}},\ \bibinfo {pages} {034001} (\bibinfo {year} {2021})},\ \Eprint
  {http://arxiv.org/abs/2006.08603} {arXiv:2006.08603 [hep-ph]} \BibitemShut
  {NoStop}%
\bibitem [{\citenamefont {Alvarez-Ruso}\ \emph {et~al.}(2020)\citenamefont
  {Alvarez-Ruso} \emph {et~al.}}]{Alvarez-Ruso:2020ezu}%
  \BibitemOpen
  \bibfield  {author} {\bibinfo {author} {\bibfnamefont {L.}~\bibnamefont
  {Alvarez-Ruso}} \emph {et~al.},\ }in\ \href@noop {} {\emph {\bibinfo
  {booktitle} {{2022 Snowmass Summer Study}}}}\ (\bibinfo {year} {2020})\
  \Eprint {http://arxiv.org/abs/2009.04285} {arXiv:2009.04285 [hep-ex]}
  \BibitemShut {NoStop}%
\bibitem [{\citenamefont {Aguilar-Arevalo}\ \emph {et~al.}(2007)\citenamefont
  {Aguilar-Arevalo} \emph {et~al.}}]{AguilarArevalo:2007it}%
  \BibitemOpen
  \bibfield  {author} {\bibinfo {author} {\bibfnamefont {A.~A.}\ \bibnamefont
  {Aguilar-Arevalo}} \emph {et~al.} (\bibinfo {collaboration} {MiniBooNE}),\
  }\href {\doibase 10.1103/PhysRevLett.98.231801} {\bibfield  {journal}
  {\bibinfo  {journal} {Phys. Rev. Lett.}\ }\textbf {\bibinfo {volume} {98}},\
  \bibinfo {pages} {231801} (\bibinfo {year} {2007})},\ \Eprint
  {http://arxiv.org/abs/0704.1500} {arXiv:0704.1500 [hep-ex]} \BibitemShut
  {NoStop}%
\bibitem [{\citenamefont {Abe}\ \emph {et~al.}(2020)\citenamefont {Abe} \emph
  {et~al.}}]{Abe:2019vii}%
  \BibitemOpen
  \bibfield  {author} {\bibinfo {author} {\bibfnamefont {K.}~\bibnamefont
  {Abe}} \emph {et~al.} (\bibinfo {collaboration} {T2K}),\ }\href {\doibase
  10.1038/s41586-020-2177-0} {\bibfield  {journal} {\bibinfo  {journal}
  {Nature}\ }\textbf {\bibinfo {volume} {580}},\ \bibinfo {pages} {339}
  (\bibinfo {year} {2020})},\ \bibinfo {note} {[Erratum: Nature 583, E16
  (2020)]},\ \Eprint {http://arxiv.org/abs/1910.03887} {arXiv:1910.03887
  [hep-ex]} \BibitemShut {NoStop}%
\bibitem [{\citenamefont {Evans}(2013)}]{Evans:2013pka}%
  \BibitemOpen
  \bibfield  {author} {\bibinfo {author} {\bibfnamefont {J.}~\bibnamefont
  {Evans}} (\bibinfo {collaboration} {MINOS}),\ }\href {\doibase
  10.1155/2013/182537} {\bibfield  {journal} {\bibinfo  {journal} {{Adv. High
  Energy Phys.}}\ }\textbf {\bibinfo {volume} {2013}},\ \bibinfo {pages}
  {182537} (\bibinfo {year} {2013})},\ \Eprint {http://arxiv.org/abs/1307.0721}
  {arXiv:1307.0721 [hep-ex]} \BibitemShut {NoStop}%
\bibitem [{\citenamefont {Ayres}\ \emph {et~al.}(2004)\citenamefont {Ayres}
  \emph {et~al.}}]{Ayres:2004js}%
  \BibitemOpen
  \bibfield  {author} {\bibinfo {author} {\bibfnamefont {D.~S.}\ \bibnamefont
  {Ayres}} \emph {et~al.} (\bibinfo {collaboration} {NOvA}),\ }\href@noop {} {\
   (\bibinfo {year} {2004})},\ \Eprint {http://arxiv.org/abs/hep-ex/0503053}
  {arXiv:hep-ex/0503053 [hep-ex]} \BibitemShut {NoStop}%
\bibitem [{\citenamefont {Bilenky}\ and\ \citenamefont
  {Pontecorvo}(1978)}]{Bilenky:1978nj}%
  \BibitemOpen
  \bibfield  {author} {\bibinfo {author} {\bibfnamefont {S.~M.}\ \bibnamefont
  {Bilenky}}\ and\ \bibinfo {author} {\bibfnamefont {B.}~\bibnamefont
  {Pontecorvo}},\ }\href {\doibase 10.1016/0370-1573(78)90095-9} {\bibfield
  {journal} {\bibinfo  {journal} {Phys. Rept.}\ }\textbf {\bibinfo {volume}
  {41}},\ \bibinfo {pages} {225} (\bibinfo {year} {1978})}\BibitemShut
  {NoStop}%
\bibitem [{\citenamefont {Bilenky}(2013)}]{Bilenky:2012qb}%
  \BibitemOpen
  \bibfield  {author} {\bibinfo {author} {\bibfnamefont {S.~M.}\ \bibnamefont
  {Bilenky}},\ }\href {\doibase 10.1140/epjh/e2012-20068-9} {\bibfield
  {journal} {\bibinfo  {journal} {Eur. Phys. J. H}\ }\textbf {\bibinfo {volume}
  {38}},\ \bibinfo {pages} {345} (\bibinfo {year} {2013})},\ \Eprint
  {http://arxiv.org/abs/1210.3065} {arXiv:1210.3065 [hep-ph]} \BibitemShut
  {NoStop}%
\bibitem [{\citenamefont {Llewellyn~Smith}(1972)}]{LlewellynSmith:1971uhs}%
  \BibitemOpen
  \bibfield  {author} {\bibinfo {author} {\bibfnamefont {C.~H.}\ \bibnamefont
  {Llewellyn~Smith}},\ }\bibfield  {booktitle} {\emph {\bibinfo {booktitle}
  {{Gauge Theories and Neutrino Physics, Jacob, 1978:0175}}},\ }\href {\doibase
  10.1016/0370-1573(72)90010-5} {\bibfield  {journal} {\bibinfo  {journal}
  {Phys. Rept.}\ }\textbf {\bibinfo {volume} {3}},\ \bibinfo {pages} {261}
  (\bibinfo {year} {1972})}\BibitemShut {NoStop}%
\bibitem [{\citenamefont {Adler}(1968)}]{Adler:1968tw}%
  \BibitemOpen
  \bibfield  {author} {\bibinfo {author} {\bibfnamefont {S.~L.}\ \bibnamefont
  {Adler}},\ }\href {\doibase 10.1016/0003-4916(68)90278-9} {\bibfield
  {journal} {\bibinfo  {journal} {Annals Phys.}\ }\textbf {\bibinfo {volume}
  {50}},\ \bibinfo {pages} {189} (\bibinfo {year} {1968})},\ \bibinfo {note}
  {[,225(1968)]}\BibitemShut {NoStop}%
\bibitem [{\citenamefont {Schreiner}\ and\ \citenamefont
  {Von~Hippel}(1973)}]{Schreiner:1973mj}%
  \BibitemOpen
  \bibfield  {author} {\bibinfo {author} {\bibfnamefont {P.~A.}\ \bibnamefont
  {Schreiner}}\ and\ \bibinfo {author} {\bibfnamefont {F.}~\bibnamefont
  {Von~Hippel}},\ }\href {\doibase 10.1016/0550-3213(73)90588-9} {\bibfield
  {journal} {\bibinfo  {journal} {Nucl. Phys.}\ }\textbf {\bibinfo {volume}
  {B58}},\ \bibinfo {pages} {333} (\bibinfo {year} {1973})}\BibitemShut
  {NoStop}%
\bibitem [{\citenamefont {Ravndal}(1973)}]{Ravndal:1973xx}%
  \BibitemOpen
  \bibfield  {author} {\bibinfo {author} {\bibfnamefont {F.}~\bibnamefont
  {Ravndal}},\ }\href {\doibase 10.1007/BF02722789} {\bibfield  {journal}
  {\bibinfo  {journal} {Nuovo Cim. A}\ }\textbf {\bibinfo {volume} {18}},\
  \bibinfo {pages} {385} (\bibinfo {year} {1973})}\BibitemShut {NoStop}%
\bibitem [{\citenamefont {Gershtein}\ \emph {et~al.}(1980)\citenamefont
  {Gershtein}, \citenamefont {Komachenko},\ and\ \citenamefont
  {Khlopov}}]{Gershtein:1980vd}%
  \BibitemOpen
  \bibfield  {author} {\bibinfo {author} {\bibfnamefont {S.~S.}\ \bibnamefont
  {Gershtein}}, \bibinfo {author} {\bibfnamefont {{\relax Yu}.~{\relax Ya}.}\
  \bibnamefont {Komachenko}}, \ and\ \bibinfo {author} {\bibfnamefont
  {M.~{\relax Yu}.}\ \bibnamefont {Khlopov}},\ }\href@noop {} {\bibfield
  {journal} {\bibinfo  {journal} {Sov. J. Nucl. Phys.}\ }\textbf {\bibinfo
  {volume} {32}},\ \bibinfo {pages} {861} (\bibinfo {year} {1980})},\ \bibinfo
  {note} {[Yad. Fiz.32(1980)]}\BibitemShut {NoStop}%
\bibitem [{\citenamefont {Rein}\ and\ \citenamefont
  {Sehgal}(1981)}]{Rein:1980wg}%
  \BibitemOpen
  \bibfield  {author} {\bibinfo {author} {\bibfnamefont {D.}~\bibnamefont
  {Rein}}\ and\ \bibinfo {author} {\bibfnamefont {L.~M.}\ \bibnamefont
  {Sehgal}},\ }\href {\doibase 10.1016/0003-4916(81)90242-6} {\bibfield
  {journal} {\bibinfo  {journal} {Annals Phys.}\ }\textbf {\bibinfo {volume}
  {133}},\ \bibinfo {pages} {79} (\bibinfo {year} {1981})}\BibitemShut
  {NoStop}%
\bibitem [{\citenamefont {Rein}(1987)}]{Rein:1987cb}%
  \BibitemOpen
  \bibfield  {author} {\bibinfo {author} {\bibfnamefont {D.}~\bibnamefont
  {Rein}},\ }\href {\doibase 10.1007/BF01561054} {\bibfield  {journal}
  {\bibinfo  {journal} {Z. Phys.}\ }\textbf {\bibinfo {volume} {C35}},\
  \bibinfo {pages} {43} (\bibinfo {year} {1987})}\BibitemShut {NoStop}%
\bibitem [{\citenamefont {Alvarez-Ruso}\ \emph {et~al.}(1998)\citenamefont
  {Alvarez-Ruso}, \citenamefont {Singh},\ and\ \citenamefont
  {Vicente~Vacas}}]{AlvarezRuso:1997jr}%
  \BibitemOpen
  \bibfield  {author} {\bibinfo {author} {\bibfnamefont {L.}~\bibnamefont
  {Alvarez-Ruso}}, \bibinfo {author} {\bibfnamefont {S.~K.}\ \bibnamefont
  {Singh}}, \ and\ \bibinfo {author} {\bibfnamefont {M.~J.}\ \bibnamefont
  {Vicente~Vacas}},\ }\href {\doibase 10.1103/PhysRevC.57.2693} {\bibfield
  {journal} {\bibinfo  {journal} {Phys. Rev. C}\ }\textbf {\bibinfo {volume}
  {57}},\ \bibinfo {pages} {2693} (\bibinfo {year} {1998})},\ \Eprint
  {http://arxiv.org/abs/nucl-th/9712058} {arXiv:nucl-th/9712058} \BibitemShut
  {NoStop}%
\bibitem [{\citenamefont {Sato}\ \emph {et~al.}(2003)\citenamefont {Sato},
  \citenamefont {Uno},\ and\ \citenamefont {Lee}}]{Sato:2003rq}%
  \BibitemOpen
  \bibfield  {author} {\bibinfo {author} {\bibfnamefont {T.}~\bibnamefont
  {Sato}}, \bibinfo {author} {\bibfnamefont {D.}~\bibnamefont {Uno}}, \ and\
  \bibinfo {author} {\bibfnamefont {T.~S.~H.}\ \bibnamefont {Lee}},\ }\href
  {\doibase 10.1103/PhysRevC.67.065201} {\bibfield  {journal} {\bibinfo
  {journal} {Phys. Rev.}\ }\textbf {\bibinfo {volume} {C67}},\ \bibinfo {pages}
  {065201} (\bibinfo {year} {2003})},\ \Eprint
  {http://arxiv.org/abs/nucl-th/0303050} {arXiv:nucl-th/0303050 [nucl-th]}
  \BibitemShut {NoStop}%
\bibitem [{\citenamefont {Kuzmin}\ \emph
  {et~al.}(2005{\natexlab{a}})\citenamefont {Kuzmin}, \citenamefont
  {Lyubushkin},\ and\ \citenamefont {Naumov}}]{Kuzmin:2004ya}%
  \BibitemOpen
  \bibfield  {author} {\bibinfo {author} {\bibfnamefont {K.~S.}\ \bibnamefont
  {Kuzmin}}, \bibinfo {author} {\bibfnamefont {V.~V.}\ \bibnamefont
  {Lyubushkin}}, \ and\ \bibinfo {author} {\bibfnamefont {V.~A.}\ \bibnamefont
  {Naumov}},\ }\href {\doibase 10.1016/j.nuclphysbps.2004.11.213} {\bibfield
  {journal} {\bibinfo  {journal} {Nucl. Phys. B Proc. Suppl.}\ }\textbf
  {\bibinfo {volume} {139}},\ \bibinfo {pages} {158} (\bibinfo {year}
  {2005}{\natexlab{a}})},\ \Eprint {http://arxiv.org/abs/hep-ph/0408106}
  {arXiv:hep-ph/0408106} \BibitemShut {NoStop}%
\bibitem [{\citenamefont {Lalakulich}\ \emph {et~al.}(2006)\citenamefont
  {Lalakulich}, \citenamefont {Paschos},\ and\ \citenamefont
  {Piranishvili}}]{Lalakulich:2006sw}%
  \BibitemOpen
  \bibfield  {author} {\bibinfo {author} {\bibfnamefont {O.}~\bibnamefont
  {Lalakulich}}, \bibinfo {author} {\bibfnamefont {E.~A.}\ \bibnamefont
  {Paschos}}, \ and\ \bibinfo {author} {\bibfnamefont {G.}~\bibnamefont
  {Piranishvili}},\ }\href {\doibase 10.1103/PhysRevD.74.014009} {\bibfield
  {journal} {\bibinfo  {journal} {Phys. Rev. D}\ }\textbf {\bibinfo {volume}
  {74}},\ \bibinfo {pages} {014009} (\bibinfo {year} {2006})},\ \Eprint
  {http://arxiv.org/abs/hep-ph/0602210} {arXiv:hep-ph/0602210} \BibitemShut
  {NoStop}%
\bibitem [{\citenamefont {Hernandez}\ \emph
  {et~al.}(2007{\natexlab{b}})\citenamefont {Hernandez}, \citenamefont
  {Nieves},\ and\ \citenamefont {Valverde}}]{Hernandez:2006yg}%
  \BibitemOpen
  \bibfield  {author} {\bibinfo {author} {\bibfnamefont {E.}~\bibnamefont
  {Hernandez}}, \bibinfo {author} {\bibfnamefont {J.}~\bibnamefont {Nieves}}, \
  and\ \bibinfo {author} {\bibfnamefont {M.}~\bibnamefont {Valverde}},\ }\href
  {\doibase 10.1016/j.physletb.2007.02.051} {\bibfield  {journal} {\bibinfo
  {journal} {Phys. Lett.}\ }\textbf {\bibinfo {volume} {B647}},\ \bibinfo
  {pages} {452} (\bibinfo {year} {2007}{\natexlab{b}})},\ \Eprint
  {http://arxiv.org/abs/hep-ph/0608119} {arXiv:hep-ph/0608119 [hep-ph]}
  \BibitemShut {NoStop}%
\bibitem [{\citenamefont {Graczyk}\ and\ \citenamefont
  {Sobczyk}(2008{\natexlab{a}})}]{Graczyk:2007bc}%
  \BibitemOpen
  \bibfield  {author} {\bibinfo {author} {\bibfnamefont {K.~M.}\ \bibnamefont
  {Graczyk}}\ and\ \bibinfo {author} {\bibfnamefont {J.~T.}\ \bibnamefont
  {Sobczyk}},\ }\href {\doibase 10.1103/PhysRevD.79.079903,
  10.1103/PhysRevD.77.053001} {\bibfield  {journal} {\bibinfo  {journal} {Phys.
  Rev.}\ }\textbf {\bibinfo {volume} {D77}},\ \bibinfo {pages} {053001}
  (\bibinfo {year} {2008}{\natexlab{a}})},\ \bibinfo {note} {[Erratum: Phys.
  Rev.D79,079903(2009)]},\ \Eprint {http://arxiv.org/abs/0707.3561}
  {arXiv:0707.3561 [hep-ph]} \BibitemShut {NoStop}%
\bibitem [{\citenamefont {Berger}\ and\ \citenamefont
  {Sehgal}(2007)}]{Berger:2007rq}%
  \BibitemOpen
  \bibfield  {author} {\bibinfo {author} {\bibfnamefont {C.}~\bibnamefont
  {Berger}}\ and\ \bibinfo {author} {\bibfnamefont {L.~M.}\ \bibnamefont
  {Sehgal}},\ }\href {\doibase 10.1103/PhysRevD.76.113004} {\bibfield
  {journal} {\bibinfo  {journal} {Phys. Rev. D}\ }\textbf {\bibinfo {volume}
  {76}},\ \bibinfo {pages} {113004} (\bibinfo {year} {2007})},\ \Eprint
  {http://arxiv.org/abs/0709.4378} {arXiv:0709.4378 [hep-ph]} \BibitemShut
  {NoStop}%
\bibitem [{\citenamefont {Barbero}\ \emph {et~al.}(2008)\citenamefont
  {Barbero}, \citenamefont {Lopez~Castro},\ and\ \citenamefont
  {Mariano}}]{Barbero:2008zza}%
  \BibitemOpen
  \bibfield  {author} {\bibinfo {author} {\bibfnamefont {C.}~\bibnamefont
  {Barbero}}, \bibinfo {author} {\bibfnamefont {G.}~\bibnamefont
  {Lopez~Castro}}, \ and\ \bibinfo {author} {\bibfnamefont {A.}~\bibnamefont
  {Mariano}},\ }\href {\doibase 10.1016/j.physletb.2008.05.011} {\bibfield
  {journal} {\bibinfo  {journal} {Phys. Lett.}\ }\textbf {\bibinfo {volume}
  {B664}},\ \bibinfo {pages} {70} (\bibinfo {year} {2008})}\BibitemShut
  {NoStop}%
\bibitem [{\citenamefont {Graczyk}\ and\ \citenamefont
  {Sobczyk}(2008{\natexlab{b}})}]{Graczyk:2008zz}%
  \BibitemOpen
  \bibfield  {author} {\bibinfo {author} {\bibfnamefont {K.~M.}\ \bibnamefont
  {Graczyk}}\ and\ \bibinfo {author} {\bibfnamefont {J.~T.}\ \bibnamefont
  {Sobczyk}},\ }\href {\doibase 10.1103/PhysRevD.77.053003} {\bibfield
  {journal} {\bibinfo  {journal} {Phys. Rev.}\ }\textbf {\bibinfo {volume}
  {D77}},\ \bibinfo {pages} {053003} (\bibinfo {year} {2008}{\natexlab{b}})},\
  \Eprint {http://arxiv.org/abs/0709.4634} {arXiv:0709.4634 [hep-ph]}
  \BibitemShut {NoStop}%
\bibitem [{\citenamefont {Leitner}\ \emph {et~al.}(2009)\citenamefont
  {Leitner}, \citenamefont {Buss}, \citenamefont {Alvarez-Ruso},\ and\
  \citenamefont {Mosel}}]{Leitner:2008ue}%
  \BibitemOpen
  \bibfield  {author} {\bibinfo {author} {\bibfnamefont {T.}~\bibnamefont
  {Leitner}}, \bibinfo {author} {\bibfnamefont {O.}~\bibnamefont {Buss}},
  \bibinfo {author} {\bibfnamefont {L.}~\bibnamefont {Alvarez-Ruso}}, \ and\
  \bibinfo {author} {\bibfnamefont {U.}~\bibnamefont {Mosel}},\ }\href
  {\doibase 10.1103/PhysRevC.79.034601} {\bibfield  {journal} {\bibinfo
  {journal} {Phys. Rev.}\ }\textbf {\bibinfo {volume} {C79}},\ \bibinfo {pages}
  {034601} (\bibinfo {year} {2009})},\ \Eprint {http://arxiv.org/abs/0812.0587}
  {arXiv:0812.0587 [nucl-th]} \BibitemShut {NoStop}%
\bibitem [{\citenamefont {Graczyk}\ \emph {et~al.}(2009)\citenamefont
  {Graczyk}, \citenamefont {Kielczewska}, \citenamefont {Przewlocki},\ and\
  \citenamefont {Sobczyk}}]{Graczyk:2009qm}%
  \BibitemOpen
  \bibfield  {author} {\bibinfo {author} {\bibfnamefont {K.~M.}\ \bibnamefont
  {Graczyk}}, \bibinfo {author} {\bibfnamefont {D.}~\bibnamefont
  {Kielczewska}}, \bibinfo {author} {\bibfnamefont {P.}~\bibnamefont
  {Przewlocki}}, \ and\ \bibinfo {author} {\bibfnamefont {J.~T.}\ \bibnamefont
  {Sobczyk}},\ }\href {\doibase 10.1103/PhysRevD.80.093001} {\bibfield
  {journal} {\bibinfo  {journal} {Phys. Rev.}\ }\textbf {\bibinfo {volume}
  {D80}},\ \bibinfo {pages} {093001} (\bibinfo {year} {2009})},\ \Eprint
  {http://arxiv.org/abs/0908.2175} {arXiv:0908.2175 [hep-ph]} \BibitemShut
  {NoStop}%
\bibitem [{\citenamefont {Lalakulich}\ \emph {et~al.}(2010)\citenamefont
  {Lalakulich}, \citenamefont {Leitner}, \citenamefont {Buss},\ and\
  \citenamefont {Mosel}}]{Lalakulich:2010ss}%
  \BibitemOpen
  \bibfield  {author} {\bibinfo {author} {\bibfnamefont {O.}~\bibnamefont
  {Lalakulich}}, \bibinfo {author} {\bibfnamefont {T.}~\bibnamefont {Leitner}},
  \bibinfo {author} {\bibfnamefont {O.}~\bibnamefont {Buss}}, \ and\ \bibinfo
  {author} {\bibfnamefont {U.}~\bibnamefont {Mosel}},\ }\href {\doibase
  10.1103/PhysRevD.82.093001} {\bibfield  {journal} {\bibinfo  {journal} {Phys.
  Rev.}\ }\textbf {\bibinfo {volume} {D82}},\ \bibinfo {pages} {093001}
  (\bibinfo {year} {2010})},\ \Eprint {http://arxiv.org/abs/1007.0925}
  {arXiv:1007.0925 [hep-ph]} \BibitemShut {NoStop}%
\bibitem [{\citenamefont {Mariano}\ \emph {et~al.}(2011)\citenamefont
  {Mariano}, \citenamefont {Barbero},\ and\ \citenamefont
  {Lopez~Castro}}]{Mariano:2011zz}%
  \BibitemOpen
  \bibfield  {author} {\bibinfo {author} {\bibfnamefont {A.}~\bibnamefont
  {Mariano}}, \bibinfo {author} {\bibfnamefont {C.}~\bibnamefont {Barbero}}, \
  and\ \bibinfo {author} {\bibfnamefont {G.}~\bibnamefont {Lopez~Castro}},\
  }\href {\doibase 10.1016/j.nuclphysa.2010.11.002} {\bibfield  {journal}
  {\bibinfo  {journal} {Nucl. Phys. A}\ }\textbf {\bibinfo {volume} {849}},\
  \bibinfo {pages} {218} (\bibinfo {year} {2011})}\BibitemShut {NoStop}%
\bibitem [{\citenamefont {Sobczyk.}\ and\ \citenamefont
  {Zmuda}(2013)}]{Sobczyk.:2012zj}%
  \BibitemOpen
  \bibfield  {author} {\bibinfo {author} {\bibfnamefont {J.~T.}\ \bibnamefont
  {Sobczyk.}}\ and\ \bibinfo {author} {\bibfnamefont {J.}~\bibnamefont
  {Zmuda}},\ }\href {\doibase 10.1103/PhysRevC.87.065503} {\bibfield  {journal}
  {\bibinfo  {journal} {Phys. Rev. C}\ }\textbf {\bibinfo {volume} {87}},\
  \bibinfo {pages} {065503} (\bibinfo {year} {2013})},\ \Eprint
  {http://arxiv.org/abs/1210.6149} {arXiv:1210.6149 [nucl-th]} \BibitemShut
  {NoStop}%
\bibitem [{\citenamefont {Serot}\ and\ \citenamefont
  {Zhang}(2012)}]{Serot:2012rd}%
  \BibitemOpen
  \bibfield  {author} {\bibinfo {author} {\bibfnamefont {B.~D.}\ \bibnamefont
  {Serot}}\ and\ \bibinfo {author} {\bibfnamefont {X.}~\bibnamefont {Zhang}},\
  }\href {\doibase 10.1103/PhysRevC.86.015501} {\bibfield  {journal} {\bibinfo
  {journal} {Phys. Rev.}\ }\textbf {\bibinfo {volume} {C86}},\ \bibinfo {pages}
  {015501} (\bibinfo {year} {2012})},\ \Eprint {http://arxiv.org/abs/1206.3812}
  {arXiv:1206.3812 [nucl-th]} \BibitemShut {NoStop}%
\bibitem [{\citenamefont {Graczyk}\ \emph {et~al.}(2014)\citenamefont
  {Graczyk}, \citenamefont {Zmuda},\ and\ \citenamefont
  {Sobczyk}}]{Graczyk:2014dpa}%
  \BibitemOpen
  \bibfield  {author} {\bibinfo {author} {\bibfnamefont {K.~M.}\ \bibnamefont
  {Graczyk}}, \bibinfo {author} {\bibfnamefont {J.}~\bibnamefont {Zmuda}}, \
  and\ \bibinfo {author} {\bibfnamefont {J.~T.}\ \bibnamefont {Sobczyk}},\
  }\href {\doibase 10.1103/PhysRevD.90.093001} {\bibfield  {journal} {\bibinfo
  {journal} {Phys. Rev.}\ }\textbf {\bibinfo {volume} {D90}},\ \bibinfo {pages}
  {093001} (\bibinfo {year} {2014})},\ \Eprint {http://arxiv.org/abs/1407.5445}
  {arXiv:1407.5445 [hep-ph]} \BibitemShut {NoStop}%
\bibitem [{\citenamefont {Rafi~Alam}\ \emph {et~al.}(2016)\citenamefont
  {Rafi~Alam}, \citenamefont {Sajjad~Athar}, \citenamefont {Chauhan},\ and\
  \citenamefont {Singh}}]{Alam:2015gaa}%
  \BibitemOpen
  \bibfield  {author} {\bibinfo {author} {\bibfnamefont {M.}~\bibnamefont
  {Rafi~Alam}}, \bibinfo {author} {\bibfnamefont {M.}~\bibnamefont
  {Sajjad~Athar}}, \bibinfo {author} {\bibfnamefont {S.}~\bibnamefont
  {Chauhan}}, \ and\ \bibinfo {author} {\bibfnamefont {S.~K.}\ \bibnamefont
  {Singh}},\ }\href {\doibase 10.1142/S0218301316500105} {\bibfield  {journal}
  {\bibinfo  {journal} {Int. J. Mod. Phys.}\ }\textbf {\bibinfo {volume}
  {E25}},\ \bibinfo {pages} {1650010} (\bibinfo {year} {2016})},\ \Eprint
  {http://arxiv.org/abs/1509.08622} {arXiv:1509.08622 [hep-ph]} \BibitemShut
  {NoStop}%
\bibitem [{\citenamefont {Alvarez-Ruso}\ \emph {et~al.}(2016)\citenamefont
  {Alvarez-Ruso}, \citenamefont {Hernandez}, \citenamefont {Nieves},\ and\
  \citenamefont {Vicente~Vacas}}]{Alvarez-Ruso:2015eva}%
  \BibitemOpen
  \bibfield  {author} {\bibinfo {author} {\bibfnamefont {L.}~\bibnamefont
  {Alvarez-Ruso}}, \bibinfo {author} {\bibfnamefont {E.}~\bibnamefont
  {Hernandez}}, \bibinfo {author} {\bibfnamefont {J.}~\bibnamefont {Nieves}}, \
  and\ \bibinfo {author} {\bibfnamefont {M.~J.}\ \bibnamefont
  {Vicente~Vacas}},\ }\href {\doibase 10.1103/PhysRevD.93.014016} {\bibfield
  {journal} {\bibinfo  {journal} {Phys. Rev.}\ }\textbf {\bibinfo {volume}
  {D93}},\ \bibinfo {pages} {014016} (\bibinfo {year} {2016})},\ \Eprint
  {http://arxiv.org/abs/1510.06266} {arXiv:1510.06266 [hep-ph]} \BibitemShut
  {NoStop}%
\bibitem [{\citenamefont {Gonzalez-Jimenez}\ \emph {et~al.}(2017)\citenamefont
  {Gonzalez-Jimenez}, \citenamefont {Jachowicz}, \citenamefont {Niewczas},
  \citenamefont {Nys}, \citenamefont {Pandey}, \citenamefont {Van~Cuyck},\ and\
  \citenamefont {Van~Dessel}}]{Gonzalez-Jimenez:2016qqq}%
  \BibitemOpen
  \bibfield  {author} {\bibinfo {author} {\bibfnamefont {R.}~\bibnamefont
  {Gonzalez-Jimenez}}, \bibinfo {author} {\bibfnamefont {N.}~\bibnamefont
  {Jachowicz}}, \bibinfo {author} {\bibfnamefont {K.}~\bibnamefont {Niewczas}},
  \bibinfo {author} {\bibfnamefont {J.}~\bibnamefont {Nys}}, \bibinfo {author}
  {\bibfnamefont {V.}~\bibnamefont {Pandey}}, \bibinfo {author} {\bibfnamefont
  {T.}~\bibnamefont {Van~Cuyck}}, \ and\ \bibinfo {author} {\bibfnamefont
  {N.}~\bibnamefont {Van~Dessel}},\ }\href {\doibase
  10.1103/PhysRevD.95.113007} {\bibfield  {journal} {\bibinfo  {journal} {Phys.
  Rev.}\ }\textbf {\bibinfo {volume} {D95}},\ \bibinfo {pages} {113007}
  (\bibinfo {year} {2017})},\ \Eprint {http://arxiv.org/abs/1612.05511}
  {arXiv:1612.05511 [nucl-th]} \BibitemShut {NoStop}%
\bibitem [{\citenamefont {Hernandez}\ and\ \citenamefont
  {Nieves}(2017)}]{Hernandez:2016yfb}%
  \BibitemOpen
  \bibfield  {author} {\bibinfo {author} {\bibfnamefont {E.}~\bibnamefont
  {Hernandez}}\ and\ \bibinfo {author} {\bibfnamefont {J.}~\bibnamefont
  {Nieves}},\ }\href {\doibase 10.1103/PhysRevD.95.053007} {\bibfield
  {journal} {\bibinfo  {journal} {Phys. Rev.}\ }\textbf {\bibinfo {volume}
  {D95}},\ \bibinfo {pages} {053007} (\bibinfo {year} {2017})},\ \Eprint
  {http://arxiv.org/abs/1612.02343} {arXiv:1612.02343 [hep-ph]} \BibitemShut
  {NoStop}%
\bibitem [{\citenamefont {Yao}\ \emph {et~al.}(2018)\citenamefont {Yao},
  \citenamefont {Alvarez-Ruso}, \citenamefont {Blin},\ and\ \citenamefont
  {Vicente~Vacas}}]{Yao:2018pzc}%
  \BibitemOpen
  \bibfield  {author} {\bibinfo {author} {\bibfnamefont {D.-L.}\ \bibnamefont
  {Yao}}, \bibinfo {author} {\bibfnamefont {L.}~\bibnamefont {Alvarez-Ruso}},
  \bibinfo {author} {\bibfnamefont {A.~N.~H.}\ \bibnamefont {Blin}}, \ and\
  \bibinfo {author} {\bibfnamefont {M.~J.}\ \bibnamefont {Vicente~Vacas}},\
  }\href@noop {} {\  (\bibinfo {year} {2018})},\ \Eprint
  {http://arxiv.org/abs/1806.09364} {arXiv:1806.09364 [hep-ph]} \BibitemShut
  {NoStop}%
\bibitem [{\citenamefont {Nakamura}\ \emph {et~al.}(2015)\citenamefont
  {Nakamura}, \citenamefont {Kamano},\ and\ \citenamefont
  {Sato}}]{Nakamura:2015rta}%
  \BibitemOpen
  \bibfield  {author} {\bibinfo {author} {\bibfnamefont {S.~X.}\ \bibnamefont
  {Nakamura}}, \bibinfo {author} {\bibfnamefont {H.}~\bibnamefont {Kamano}}, \
  and\ \bibinfo {author} {\bibfnamefont {T.}~\bibnamefont {Sato}},\ }\href
  {\doibase 10.1103/PhysRevD.92.074024} {\bibfield  {journal} {\bibinfo
  {journal} {Phys. Rev.}\ }\textbf {\bibinfo {volume} {D92}},\ \bibinfo {pages}
  {074024} (\bibinfo {year} {2015})},\ \Eprint
  {http://arxiv.org/abs/1506.03403} {arXiv:1506.03403 [hep-ph]} \BibitemShut
  {NoStop}%
\bibitem [{\citenamefont {Nakamura}\ \emph {et~al.}(2019)\citenamefont
  {Nakamura}, \citenamefont {Kamano},\ and\ \citenamefont
  {Sato}}]{Nakamura:2018ntd}%
  \BibitemOpen
  \bibfield  {author} {\bibinfo {author} {\bibfnamefont {S.~X.}\ \bibnamefont
  {Nakamura}}, \bibinfo {author} {\bibfnamefont {H.}~\bibnamefont {Kamano}}, \
  and\ \bibinfo {author} {\bibfnamefont {T.}~\bibnamefont {Sato}},\ }\href
  {\doibase 10.1103/PhysRevD.99.031301} {\bibfield  {journal} {\bibinfo
  {journal} {Phys. Rev. D}\ }\textbf {\bibinfo {volume} {99}},\ \bibinfo
  {pages} {031301} (\bibinfo {year} {2019})},\ \Eprint
  {http://arxiv.org/abs/1812.00144} {arXiv:1812.00144 [hep-ph]} \BibitemShut
  {NoStop}%
\bibitem [{\citenamefont {Sobczyk}\ \emph {et~al.}(2018)\citenamefont
  {Sobczyk}, \citenamefont {Hern\'andez}, \citenamefont {Nakamura},
  \citenamefont {Nieves},\ and\ \citenamefont {Sato}}]{Sobczyk:2018ghy}%
  \BibitemOpen
  \bibfield  {author} {\bibinfo {author} {\bibfnamefont {J.~E.}\ \bibnamefont
  {Sobczyk}}, \bibinfo {author} {\bibfnamefont {E.}~\bibnamefont
  {Hern\'andez}}, \bibinfo {author} {\bibfnamefont {S.~X.}\ \bibnamefont
  {Nakamura}}, \bibinfo {author} {\bibfnamefont {J.}~\bibnamefont {Nieves}}, \
  and\ \bibinfo {author} {\bibfnamefont {T.}~\bibnamefont {Sato}},\ }\href
  {\doibase 10.1103/PhysRevD.98.073001} {\bibfield  {journal} {\bibinfo
  {journal} {Phys. Rev. D}\ }\textbf {\bibinfo {volume} {98}},\ \bibinfo
  {pages} {073001} (\bibinfo {year} {2018})},\ \Eprint
  {http://arxiv.org/abs/1807.11281} {arXiv:1807.11281 [hep-ph]} \BibitemShut
  {NoStop}%
\bibitem [{\citenamefont {Yao}\ \emph {et~al.}(2019)\citenamefont {Yao},
  \citenamefont {Alvarez-Ruso},\ and\ \citenamefont
  {Vicente~Vacas}}]{Yao:2019avf}%
  \BibitemOpen
  \bibfield  {author} {\bibinfo {author} {\bibfnamefont {D.-L.}\ \bibnamefont
  {Yao}}, \bibinfo {author} {\bibfnamefont {L.}~\bibnamefont {Alvarez-Ruso}}, \
  and\ \bibinfo {author} {\bibfnamefont {M.~J.}\ \bibnamefont
  {Vicente~Vacas}},\ }\href@noop {} {\  (\bibinfo {year} {2019})},\ \Eprint
  {http://arxiv.org/abs/1901.00773} {arXiv:1901.00773 [hep-ph]} \BibitemShut
  {NoStop}%
\bibitem [{\citenamefont {Rocco}\ \emph {et~al.}(2019)\citenamefont {Rocco},
  \citenamefont {Nakamura}, \citenamefont {Lee},\ and\ \citenamefont
  {Lovato}}]{Rocco:2019gfb}%
  \BibitemOpen
  \bibfield  {author} {\bibinfo {author} {\bibfnamefont {N.}~\bibnamefont
  {Rocco}}, \bibinfo {author} {\bibfnamefont {S.~X.}\ \bibnamefont {Nakamura}},
  \bibinfo {author} {\bibfnamefont {T.~S.~H.}\ \bibnamefont {Lee}}, \ and\
  \bibinfo {author} {\bibfnamefont {A.}~\bibnamefont {Lovato}},\ }\href
  {\doibase 10.1103/PhysRevC.100.045503} {\bibfield  {journal} {\bibinfo
  {journal} {Phys. Rev. C}\ }\textbf {\bibinfo {volume} {100}},\ \bibinfo
  {pages} {045503} (\bibinfo {year} {2019})},\ \Eprint
  {http://arxiv.org/abs/1907.01093} {arXiv:1907.01093 [nucl-th]} \BibitemShut
  {NoStop}%
\bibitem [{\citenamefont {Radecky}\ \emph {et~al.}(1982)\citenamefont
  {Radecky}, \citenamefont {Barnes}, \citenamefont {Carmony}, \citenamefont
  {Garfinkel}, \citenamefont {Derrick}, \citenamefont {Fernandez},
  \citenamefont {Hyman},\ and\ \citenamefont {Levman~{\it et
  al.}}}]{Radecky:1981fn}%
  \BibitemOpen
  \bibfield  {author} {\bibinfo {author} {\bibfnamefont {G.~M.}\ \bibnamefont
  {Radecky}}, \bibinfo {author} {\bibfnamefont {V.~E.}\ \bibnamefont {Barnes}},
  \bibinfo {author} {\bibfnamefont {D.~D.}\ \bibnamefont {Carmony}}, \bibinfo
  {author} {\bibfnamefont {A.~F.}\ \bibnamefont {Garfinkel}}, \bibinfo {author}
  {\bibfnamefont {M.}~\bibnamefont {Derrick}}, \bibinfo {author} {\bibfnamefont
  {E.}~\bibnamefont {Fernandez}}, \bibinfo {author} {\bibfnamefont
  {L.}~\bibnamefont {Hyman}}, \ and\ \bibinfo {author} {\bibfnamefont
  {G.}~\bibnamefont {Levman~{\it et al.}}},\ }\href@noop {} {\bibfield
  {journal} {\bibinfo  {journal} {Phys.\ Rev.\ D}\ }\textbf {\bibinfo {volume}
  {{\bf 26}}},\ \bibinfo {pages} {3297} (\bibinfo {year} {1982})},\ \bibinfo
  {note} {[Erratum-ibid.\ D {\bf 26} (1982) 3297]}\BibitemShut {NoStop}%
\bibitem [{\citenamefont {Kitagaki}\ \emph {et~al.}(1986)\citenamefont
  {Kitagaki}, \citenamefont {Yuta}, \citenamefont {Tanaka}, \citenamefont
  {Yamaguchi}, \citenamefont {Abe} \emph {et~al.}}]{Kitagaki:1986ct}%
  \BibitemOpen
  \bibfield  {author} {\bibinfo {author} {\bibfnamefont {T.}~\bibnamefont
  {Kitagaki}}, \bibinfo {author} {\bibfnamefont {H.}~\bibnamefont {Yuta}},
  \bibinfo {author} {\bibfnamefont {S.}~\bibnamefont {Tanaka}}, \bibinfo
  {author} {\bibfnamefont {A.}~\bibnamefont {Yamaguchi}}, \bibinfo {author}
  {\bibfnamefont {K.}~\bibnamefont {Abe}},  \emph {et~al.},\ }\href {\doibase
  10.1103/PhysRevD.34.2554} {\bibfield  {journal} {\bibinfo  {journal} {Phys.
  Rev. D}\ }\textbf {\bibinfo {volume} {34}},\ \bibinfo {pages} {2554}
  (\bibinfo {year} {1986})}\BibitemShut {NoStop}%
\bibitem [{\citenamefont {Singh}\ and\ \citenamefont
  {Arenhovel}(1986)}]{Singh:1986xh}%
  \BibitemOpen
  \bibfield  {author} {\bibinfo {author} {\bibfnamefont {S.~K.}\ \bibnamefont
  {Singh}}\ and\ \bibinfo {author} {\bibfnamefont {H.}~\bibnamefont
  {Arenhovel}},\ }\href {\doibase 10.1007/BF01294589} {\bibfield  {journal}
  {\bibinfo  {journal} {Z. Phys. A}\ }\textbf {\bibinfo {volume} {324}},\
  \bibinfo {pages} {347} (\bibinfo {year} {1986})}\BibitemShut {NoStop}%
\bibitem [{\citenamefont {Hernandez}\ \emph {et~al.}(2010)\citenamefont
  {Hernandez}, \citenamefont {Nieves}, \citenamefont {Valverde},\ and\
  \citenamefont {Vicente~Vacas}}]{Hernandez:2010bx}%
  \BibitemOpen
  \bibfield  {author} {\bibinfo {author} {\bibfnamefont {E.}~\bibnamefont
  {Hernandez}}, \bibinfo {author} {\bibfnamefont {J.}~\bibnamefont {Nieves}},
  \bibinfo {author} {\bibfnamefont {M.}~\bibnamefont {Valverde}}, \ and\
  \bibinfo {author} {\bibfnamefont {M.~J.}\ \bibnamefont {Vicente~Vacas}},\
  }\href {\doibase 10.1103/PhysRevD.81.085046} {\bibfield  {journal} {\bibinfo
  {journal} {Phys. Rev.}\ }\textbf {\bibinfo {volume} {D81}},\ \bibinfo {pages}
  {085046} (\bibinfo {year} {2010})},\ \Eprint {http://arxiv.org/abs/1001.4416}
  {arXiv:1001.4416 [hep-ph]} \BibitemShut {NoStop}%
\bibitem [{\citenamefont {Wilkinson}\ \emph {et~al.}(2014)\citenamefont
  {Wilkinson}, \citenamefont {Rodrigues}, \citenamefont {Cartwright},
  \citenamefont {Thompson},\ and\ \citenamefont
  {McFarland}}]{Wilkinson:2014yfa}%
  \BibitemOpen
  \bibfield  {author} {\bibinfo {author} {\bibfnamefont {C.}~\bibnamefont
  {Wilkinson}}, \bibinfo {author} {\bibfnamefont {P.}~\bibnamefont
  {Rodrigues}}, \bibinfo {author} {\bibfnamefont {S.}~\bibnamefont
  {Cartwright}}, \bibinfo {author} {\bibfnamefont {L.}~\bibnamefont
  {Thompson}}, \ and\ \bibinfo {author} {\bibfnamefont {K.}~\bibnamefont
  {McFarland}},\ }\href {\doibase 10.1103/PhysRevD.90.112017} {\bibfield
  {journal} {\bibinfo  {journal} {Phys. Rev. D}\ }\textbf {\bibinfo {volume}
  {90}},\ \bibinfo {pages} {112017} (\bibinfo {year} {2014})},\ \Eprint
  {http://arxiv.org/abs/1411.4482} {arXiv:1411.4482 [hep-ex]} \BibitemShut
  {NoStop}%
\bibitem [{\citenamefont {Adler}(1963)}]{Adler1963}%
  \BibitemOpen
  \bibfield  {author} {\bibinfo {author} {\bibfnamefont {S.~L.}\ \bibnamefont
  {Adler}},\ }\href {\doibase 10.1007/BF02828811} {\bibfield  {journal}
  {\bibinfo  {journal} {Il Nuovo Cimento (1955-1965)}\ }\textbf {\bibinfo
  {volume} {30}},\ \bibinfo {pages} {1020} (\bibinfo {year}
  {1963})}\BibitemShut {NoStop}%
\bibitem [{\citenamefont {Florescu}\ and\ \citenamefont
  {Minnaert}(1968)}]{PhysRev.168.1662}%
  \BibitemOpen
  \bibfield  {author} {\bibinfo {author} {\bibfnamefont {V.}~\bibnamefont
  {Florescu}}\ and\ \bibinfo {author} {\bibfnamefont {P.}~\bibnamefont
  {Minnaert}},\ }\href {\doibase 10.1103/PhysRev.168.1662} {\bibfield
  {journal} {\bibinfo  {journal} {Phys. Rev.}\ }\textbf {\bibinfo {volume}
  {168}},\ \bibinfo {pages} {1662} (\bibinfo {year} {1968})}\BibitemShut
  {NoStop}%
\bibitem [{\citenamefont {Pais}(1971)}]{Pais:1971er}%
  \BibitemOpen
  \bibfield  {author} {\bibinfo {author} {\bibfnamefont {A.}~\bibnamefont
  {Pais}},\ }\href {\doibase 10.1016/0003-4916(71)90018-2} {\bibfield
  {journal} {\bibinfo  {journal} {Annals Phys.}\ }\textbf {\bibinfo {volume}
  {63}},\ \bibinfo {pages} {361} (\bibinfo {year} {1971})}\BibitemShut
  {NoStop}%
\bibitem [{\citenamefont {Block}(1965)}]{Block:1965zol}%
  \BibitemOpen
  \bibfield  {author} {\bibinfo {author} {\bibfnamefont {M.~M.}\ \bibnamefont
  {Block}},\ }in\ \href {\doibase 10.1016/B978-1-4832-5648-1.50017-8} {\emph
  {\bibinfo {booktitle} {{Symmetries in Elementary Particle Physics:
  International School of Physics Ettore Majorana, Erice, Italy, Aug 1964}}}}\
  (\bibinfo {year} {1965})\ p.\ \bibinfo {pages} {341}\BibitemShut {NoStop}%
\bibitem [{\citenamefont {Lee}\ and\ \citenamefont
  {Yang}(1962)}]{PhysRev.126.2239}%
  \BibitemOpen
  \bibfield  {author} {\bibinfo {author} {\bibfnamefont {T.~D.}\ \bibnamefont
  {Lee}}\ and\ \bibinfo {author} {\bibfnamefont {C.~N.}\ \bibnamefont {Yang}},\
  }\href {\doibase 10.1103/PhysRev.126.2239} {\bibfield  {journal} {\bibinfo
  {journal} {Phys. Rev.}\ }\textbf {\bibinfo {volume} {126}},\ \bibinfo {pages}
  {2239} (\bibinfo {year} {1962})}\BibitemShut {NoStop}%
\bibitem [{\citenamefont {Cheng}\ and\ \citenamefont
  {Tung}(1971)}]{PhysRevD.3.733}%
  \BibitemOpen
  \bibfield  {author} {\bibinfo {author} {\bibfnamefont {T.~P.}\ \bibnamefont
  {Cheng}}\ and\ \bibinfo {author} {\bibfnamefont {W.-K.}\ \bibnamefont
  {Tung}},\ }\href {\doibase 10.1103/PhysRevD.3.733} {\bibfield  {journal}
  {\bibinfo  {journal} {Phys. Rev. D}\ }\textbf {\bibinfo {volume} {3}},\
  \bibinfo {pages} {733} (\bibinfo {year} {1971})}\BibitemShut {NoStop}%
\bibitem [{\citenamefont {Tarrach}(1974)}]{TARRACH197470}%
  \BibitemOpen
  \bibfield  {author} {\bibinfo {author} {\bibfnamefont {R.}~\bibnamefont
  {Tarrach}},\ }\href {\doibase https://doi.org/10.1016/0550-3213(74)90358-7}
  {\bibfield  {journal} {\bibinfo  {journal} {Nuclear Physics B}\ }\textbf
  {\bibinfo {volume} {70}},\ \bibinfo {pages} {70} (\bibinfo {year}
  {1974})}\BibitemShut {NoStop}%
\bibitem [{\citenamefont {Oliver}\ and\ \citenamefont
  {Pham}(1974)}]{PhysRevD.10.993}%
  \BibitemOpen
  \bibfield  {author} {\bibinfo {author} {\bibfnamefont {L.}~\bibnamefont
  {Oliver}}\ and\ \bibinfo {author} {\bibfnamefont {T.~N.}\ \bibnamefont
  {Pham}},\ }\href {\doibase 10.1103/PhysRevD.10.993} {\bibfield  {journal}
  {\bibinfo  {journal} {Phys. Rev. D}\ }\textbf {\bibinfo {volume} {10}},\
  \bibinfo {pages} {993} (\bibinfo {year} {1974})}\BibitemShut {NoStop}%
\bibitem [{\citenamefont {Kim}\ \emph {et~al.}(1978)\citenamefont {Kim},
  \citenamefont {Langacker},\ and\ \citenamefont {Sarkar}}]{PhysRevD.18.123}%
  \BibitemOpen
  \bibfield  {author} {\bibinfo {author} {\bibfnamefont {J.~E.}\ \bibnamefont
  {Kim}}, \bibinfo {author} {\bibfnamefont {P.}~\bibnamefont {Langacker}}, \
  and\ \bibinfo {author} {\bibfnamefont {S.}~\bibnamefont {Sarkar}},\ }\href
  {\doibase 10.1103/PhysRevD.18.123} {\bibfield  {journal} {\bibinfo  {journal}
  {Phys. Rev. D}\ }\textbf {\bibinfo {volume} {18}},\ \bibinfo {pages} {123}
  (\bibinfo {year} {1978})}\BibitemShut {NoStop}%
\bibitem [{\citenamefont {Ridener}\ and\ \citenamefont
  {Good}(1983)}]{PhysRevD.28.2875}%
  \BibitemOpen
  \bibfield  {author} {\bibinfo {author} {\bibfnamefont {F.~L.}\ \bibnamefont
  {Ridener}}\ and\ \bibinfo {author} {\bibfnamefont {R.~H.}\ \bibnamefont
  {Good}},\ }\href {\doibase 10.1103/PhysRevD.28.2875} {\bibfield  {journal}
  {\bibinfo  {journal} {Phys. Rev. D}\ }\textbf {\bibinfo {volume} {28}},\
  \bibinfo {pages} {2875} (\bibinfo {year} {1983})}\BibitemShut {NoStop}%
\bibitem [{\citenamefont {Ridener}\ \emph {et~al.}(1985)\citenamefont
  {Ridener}, \citenamefont {Song},\ and\ \citenamefont
  {Good}}]{PhysRevD.32.2921}%
  \BibitemOpen
  \bibfield  {author} {\bibinfo {author} {\bibfnamefont {F.~L.}\ \bibnamefont
  {Ridener}}, \bibinfo {author} {\bibfnamefont {H.~S.}\ \bibnamefont {Song}}, \
  and\ \bibinfo {author} {\bibfnamefont {R.~H.}\ \bibnamefont {Good}},\ }\href
  {\doibase 10.1103/PhysRevD.32.2921} {\bibfield  {journal} {\bibinfo
  {journal} {Phys. Rev. D}\ }\textbf {\bibinfo {volume} {32}},\ \bibinfo
  {pages} {2921} (\bibinfo {year} {1985})}\BibitemShut {NoStop}%
\bibitem [{\citenamefont {Jachowicz}\ \emph {et~al.}(2004)\citenamefont
  {Jachowicz}, \citenamefont {Vantournhout}, \citenamefont {Ryckebusch},\ and\
  \citenamefont {Heyde}}]{Jachowicz:2004we}%
  \BibitemOpen
  \bibfield  {author} {\bibinfo {author} {\bibfnamefont {N.}~\bibnamefont
  {Jachowicz}}, \bibinfo {author} {\bibfnamefont {K.}~\bibnamefont
  {Vantournhout}}, \bibinfo {author} {\bibfnamefont {J.}~\bibnamefont
  {Ryckebusch}}, \ and\ \bibinfo {author} {\bibfnamefont {K.}~\bibnamefont
  {Heyde}},\ }\href {\doibase 10.1103/PhysRevLett.93.082501} {\bibfield
  {journal} {\bibinfo  {journal} {Phys. Rev. Lett.}\ }\textbf {\bibinfo
  {volume} {93}},\ \bibinfo {pages} {082501} (\bibinfo {year}
  {2004})}\BibitemShut {NoStop}%
\bibitem [{\citenamefont {Jachowicz}\ \emph {et~al.}(2005)\citenamefont
  {Jachowicz}, \citenamefont {Vantournhout}, \citenamefont {Ryckebusch},\ and\
  \citenamefont {Heyde}}]{PhysRevC.71.034604}%
  \BibitemOpen
  \bibfield  {author} {\bibinfo {author} {\bibfnamefont {N.}~\bibnamefont
  {Jachowicz}}, \bibinfo {author} {\bibfnamefont {K.}~\bibnamefont
  {Vantournhout}}, \bibinfo {author} {\bibfnamefont {J.}~\bibnamefont
  {Ryckebusch}}, \ and\ \bibinfo {author} {\bibfnamefont {K.}~\bibnamefont
  {Heyde}},\ }\href {\doibase 10.1103/PhysRevC.71.034604} {\bibfield  {journal}
  {\bibinfo  {journal} {Phys. Rev. C}\ }\textbf {\bibinfo {volume} {71}},\
  \bibinfo {pages} {034604} (\bibinfo {year} {2005})}\BibitemShut {NoStop}%
\bibitem [{\citenamefont {Hagiwara}\ \emph {et~al.}(2003)\citenamefont
  {Hagiwara}, \citenamefont {Mawatari},\ and\ \citenamefont
  {Yokoya}}]{Hagiwara:2003di}%
  \BibitemOpen
  \bibfield  {author} {\bibinfo {author} {\bibfnamefont {K.}~\bibnamefont
  {Hagiwara}}, \bibinfo {author} {\bibfnamefont {K.}~\bibnamefont {Mawatari}},
  \ and\ \bibinfo {author} {\bibfnamefont {H.}~\bibnamefont {Yokoya}},\ }\href
  {\doibase 10.1016/S0550-3213(03)00575-3} {\bibfield  {journal} {\bibinfo
  {journal} {Nucl. Phys.}\ }\textbf {\bibinfo {volume} {B668}},\ \bibinfo
  {pages} {364} (\bibinfo {year} {2003})},\ \bibinfo {note} {[Erratum: Nucl.
  Phys.B701,405(2004)]},\ \Eprint {http://arxiv.org/abs/hep-ph/0305324}
  {arXiv:hep-ph/0305324 [hep-ph]} \BibitemShut {NoStop}%
\bibitem [{\citenamefont {Kuzmin}\ \emph
  {et~al.}(2005{\natexlab{b}})\citenamefont {Kuzmin}, \citenamefont
  {Lyubushkin},\ and\ \citenamefont {Naumov}}]{Kuzmin:2004yb}%
  \BibitemOpen
  \bibfield  {author} {\bibinfo {author} {\bibfnamefont {K.~S.}\ \bibnamefont
  {Kuzmin}}, \bibinfo {author} {\bibfnamefont {V.~V.}\ \bibnamefont
  {Lyubushkin}}, \ and\ \bibinfo {author} {\bibfnamefont {V.~A.}\ \bibnamefont
  {Naumov}},\ }\bibfield  {booktitle} {\emph {\bibinfo {booktitle}
  {{Proceedings, 3rd International Workshop on Neutrino-nucleus interactions in
  the few GeV region (NUINT 04): Assergi, Italy, March 17-21, 2004}}},\ }\href
  {\doibase 10.1016/j.nuclphysbps.2004.11.221} {\bibfield  {journal} {\bibinfo
  {journal} {Nucl. Phys. Proc. Suppl.}\ }\textbf {\bibinfo {volume} {139}},\
  \bibinfo {pages} {154} (\bibinfo {year} {2005}{\natexlab{b}})},\ \bibinfo
  {note} {[,154(2004)]},\ \Eprint {http://arxiv.org/abs/hep-ph/0408107}
  {arXiv:hep-ph/0408107 [hep-ph]} \BibitemShut {NoStop}%
\bibitem [{\citenamefont {Kuzmin}\ \emph {et~al.}(2004)\citenamefont {Kuzmin},
  \citenamefont {Lyubushkin},\ and\ \citenamefont {Naumov}}]{Kuzmin:2003ji}%
  \BibitemOpen
  \bibfield  {author} {\bibinfo {author} {\bibfnamefont {K.~S.}\ \bibnamefont
  {Kuzmin}}, \bibinfo {author} {\bibfnamefont {V.~V.}\ \bibnamefont
  {Lyubushkin}}, \ and\ \bibinfo {author} {\bibfnamefont {V.~A.}\ \bibnamefont
  {Naumov}},\ }\bibfield  {booktitle} {\emph {\bibinfo {booktitle}
  {{Proceedings, 10th Advanced Research Workshop on High-Energy Spin Physics
  (SPIN-03): Dubna, Russia, September 16-20, 2003}}},\ }\href {\doibase
  10.1142/S0217732304016172} {\bibfield  {journal} {\bibinfo  {journal} {Mod.
  Phys. Lett.}\ }\textbf {\bibinfo {volume} {A19}},\ \bibinfo {pages} {2815}
  (\bibinfo {year} {2004})},\ \bibinfo {note} {[,125(2003)]},\ \Eprint
  {http://arxiv.org/abs/hep-ph/0312107} {arXiv:hep-ph/0312107 [hep-ph]}
  \BibitemShut {NoStop}%
\bibitem [{\citenamefont {Graczyk}(2005{\natexlab{a}})}]{Graczyk:2004vg}%
  \BibitemOpen
  \bibfield  {author} {\bibinfo {author} {\bibfnamefont {K.~M.}\ \bibnamefont
  {Graczyk}},\ }\href {\doibase 10.1016/j.nuclphysbps.2004.11.230} {\bibfield
  {journal} {\bibinfo  {journal} {Nucl. Phys. B Proc. Suppl.}\ }\textbf
  {\bibinfo {volume} {139}},\ \bibinfo {pages} {150} (\bibinfo {year}
  {2005}{\natexlab{a}})},\ \Eprint {http://arxiv.org/abs/hep-ph/0407283}
  {arXiv:hep-ph/0407283} \BibitemShut {NoStop}%
\bibitem [{\citenamefont {Graczyk}(2005{\natexlab{b}})}]{Graczyk:2004uy}%
  \BibitemOpen
  \bibfield  {author} {\bibinfo {author} {\bibfnamefont {K.~M.}\ \bibnamefont
  {Graczyk}},\ }\href {\doibase 10.1016/j.nuclphysa.2004.10.029} {\bibfield
  {journal} {\bibinfo  {journal} {Nucl. Phys.}\ }\textbf {\bibinfo {volume}
  {A748}},\ \bibinfo {pages} {313} (\bibinfo {year} {2005}{\natexlab{b}})},\
  \Eprint {http://arxiv.org/abs/hep-ph/0407275} {arXiv:hep-ph/0407275 [hep-ph]}
  \BibitemShut {NoStop}%
\bibitem [{\citenamefont {Levy}(2009)}]{Levy:2004rk}%
  \BibitemOpen
  \bibfield  {author} {\bibinfo {author} {\bibfnamefont {J.-M.}\ \bibnamefont
  {Levy}},\ }\href {\doibase 10.1088/0954-3899/36/5/055002} {\bibfield
  {journal} {\bibinfo  {journal} {J. Phys. G}\ }\textbf {\bibinfo {volume}
  {36}},\ \bibinfo {pages} {055002} (\bibinfo {year} {2009})},\ \Eprint
  {http://arxiv.org/abs/hep-ph/0407371} {arXiv:hep-ph/0407371} \BibitemShut
  {NoStop}%
\bibitem [{\citenamefont {Bourrely}\ \emph {et~al.}(2004)\citenamefont
  {Bourrely}, \citenamefont {Soffer},\ and\ \citenamefont
  {Teryaev}}]{PhysRevD.69.114019}%
  \BibitemOpen
  \bibfield  {author} {\bibinfo {author} {\bibfnamefont {C.}~\bibnamefont
  {Bourrely}}, \bibinfo {author} {\bibfnamefont {J.}~\bibnamefont {Soffer}}, \
  and\ \bibinfo {author} {\bibfnamefont {O.~V.}\ \bibnamefont {Teryaev}},\
  }\href {\doibase 10.1103/PhysRevD.69.114019} {\bibfield  {journal} {\bibinfo
  {journal} {Phys. Rev. D}\ }\textbf {\bibinfo {volume} {69}},\ \bibinfo
  {pages} {114019} (\bibinfo {year} {2004})}\BibitemShut {NoStop}%
\bibitem [{\citenamefont {Lagoda}\ \emph {et~al.}(2007)\citenamefont {Lagoda}
  \emph {et~al.}}]{Lagoda:2007zz}%
  \BibitemOpen
  \bibfield  {author} {\bibinfo {author} {\bibfnamefont {J.}~\bibnamefont
  {Lagoda}} \emph {et~al.},\ }\href@noop {} {\bibfield  {journal} {\bibinfo
  {journal} {Acta Phys. Polon. B}\ }\textbf {\bibinfo {volume} {38}},\ \bibinfo
  {pages} {2083} (\bibinfo {year} {2007})}\BibitemShut {NoStop}%
\bibitem [{\citenamefont {Valverde}\ \emph {et~al.}(2006)\citenamefont
  {Valverde}, \citenamefont {Amaro}, \citenamefont {Nieves},\ and\
  \citenamefont {Maieron}}]{Valverde:2006yi}%
  \BibitemOpen
  \bibfield  {author} {\bibinfo {author} {\bibfnamefont {M.}~\bibnamefont
  {Valverde}}, \bibinfo {author} {\bibfnamefont {J.~E.}\ \bibnamefont {Amaro}},
  \bibinfo {author} {\bibfnamefont {J.}~\bibnamefont {Nieves}}, \ and\ \bibinfo
  {author} {\bibfnamefont {C.}~\bibnamefont {Maieron}},\ }\href {\doibase
  10.1016/j.physletb.2006.08.087} {\bibfield  {journal} {\bibinfo  {journal}
  {Phys. Lett.}\ }\textbf {\bibinfo {volume} {B642}},\ \bibinfo {pages} {218}
  (\bibinfo {year} {2006})}\BibitemShut {NoStop}%
\bibitem [{\citenamefont {Meucci}\ \emph {et~al.}(2008)\citenamefont {Meucci},
  \citenamefont {Giusti},\ and\ \citenamefont {Pacati}}]{PhysRevC.77.034606}%
  \BibitemOpen
  \bibfield  {author} {\bibinfo {author} {\bibfnamefont {A.}~\bibnamefont
  {Meucci}}, \bibinfo {author} {\bibfnamefont {C.}~\bibnamefont {Giusti}}, \
  and\ \bibinfo {author} {\bibfnamefont {F.~D.}\ \bibnamefont {Pacati}},\
  }\href {\doibase 10.1103/PhysRevC.77.034606} {\bibfield  {journal} {\bibinfo
  {journal} {Phys. Rev. C}\ }\textbf {\bibinfo {volume} {77}},\ \bibinfo
  {pages} {034606} (\bibinfo {year} {2008})}\BibitemShut {NoStop}%
\bibitem [{\citenamefont {Bilenky}\ and\ \citenamefont
  {Christova}(2013{\natexlab{a}})}]{Bilenky:2013iua}%
  \BibitemOpen
  \bibfield  {author} {\bibinfo {author} {\bibfnamefont {S.~M.}\ \bibnamefont
  {Bilenky}}\ and\ \bibinfo {author} {\bibfnamefont {E.}~\bibnamefont
  {Christova}},\ }\href {\doibase 10.1134/S154747711307011X} {\bibfield
  {journal} {\bibinfo  {journal} {Phys. Part. Nucl. Lett.}\ }\textbf {\bibinfo
  {volume} {10}},\ \bibinfo {pages} {651} (\bibinfo {year}
  {2013}{\natexlab{a}})},\ \Eprint {http://arxiv.org/abs/1307.7275}
  {arXiv:1307.7275 [hep-ph]} \BibitemShut {NoStop}%
\bibitem [{\citenamefont {Bilenky}\ and\ \citenamefont
  {Christova}(2013{\natexlab{b}})}]{Bilenky:2013fra}%
  \BibitemOpen
  \bibfield  {author} {\bibinfo {author} {\bibfnamefont {S.~M.}\ \bibnamefont
  {Bilenky}}\ and\ \bibinfo {author} {\bibfnamefont {E.}~\bibnamefont
  {Christova}},\ }\href {\doibase 10.1088/0954-3899/40/7/075004} {\bibfield
  {journal} {\bibinfo  {journal} {J. Phys.}\ }\textbf {\bibinfo {volume}
  {G40}},\ \bibinfo {pages} {075004} (\bibinfo {year} {2013}{\natexlab{b}})},\
  \Eprint {http://arxiv.org/abs/1303.3710} {arXiv:1303.3710 [hep-ph]}
  \BibitemShut {NoStop}%
\bibitem [{\citenamefont {Graczyk}\ and\ \citenamefont
  {Kowal}(2017)}]{Graczyk:2017ngi}%
  \BibitemOpen
  \bibfield  {author} {\bibinfo {author} {\bibfnamefont {K.~M.}\ \bibnamefont
  {Graczyk}}\ and\ \bibinfo {author} {\bibfnamefont {B.~E.}\ \bibnamefont
  {Kowal}},\ }\bibfield  {booktitle} {\emph {\bibinfo {booktitle}
  {{Proceedings, 41st International Conference of Theoretical Physics: Matter
  to the Deepest: Kroczyce, Poland, September 4-8, 2017}}},\ }\href {\doibase
  10.5506/APhysPolB.48.2219} {\bibfield  {journal} {\bibinfo  {journal} {Acta
  Phys. Polon.}\ }\textbf {\bibinfo {volume} {B48}},\ \bibinfo {pages} {2219}
  (\bibinfo {year} {2017})}\BibitemShut {NoStop}%
\bibitem [{\citenamefont {Fatima}\ \emph {et~al.}(2018)\citenamefont {Fatima},
  \citenamefont {Sajjad~Athar},\ and\ \citenamefont {Singh}}]{Fatima:2018tzs}%
  \BibitemOpen
  \bibfield  {author} {\bibinfo {author} {\bibfnamefont {A.}~\bibnamefont
  {Fatima}}, \bibinfo {author} {\bibfnamefont {M.}~\bibnamefont
  {Sajjad~Athar}}, \ and\ \bibinfo {author} {\bibfnamefont {S.~K.}\
  \bibnamefont {Singh}},\ }\href {\doibase 10.1103/PhysRevD.98.033005}
  {\bibfield  {journal} {\bibinfo  {journal} {Phys. Rev.}\ }\textbf {\bibinfo
  {volume} {D98}},\ \bibinfo {pages} {033005} (\bibinfo {year} {2018})},\
  \Eprint {http://arxiv.org/abs/1806.08597} {arXiv:1806.08597 [hep-ph]}
  \BibitemShut {NoStop}%
\bibitem [{\citenamefont {Sobczyk}\ \emph {et~al.}(2019)\citenamefont
  {Sobczyk}, \citenamefont {Rocco},\ and\ \citenamefont
  {Nieves}}]{Sobczyk:2019urm}%
  \BibitemOpen
  \bibfield  {author} {\bibinfo {author} {\bibfnamefont {J.~E.}\ \bibnamefont
  {Sobczyk}}, \bibinfo {author} {\bibfnamefont {N.}~\bibnamefont {Rocco}}, \
  and\ \bibinfo {author} {\bibfnamefont {J.}~\bibnamefont {Nieves}},\ }\href
  {\doibase 10.1103/PhysRevC.100.035501} {\bibfield  {journal} {\bibinfo
  {journal} {Phys. Rev.}\ }\textbf {\bibinfo {volume} {C100}},\ \bibinfo
  {pages} {035501} (\bibinfo {year} {2019})}\BibitemShut {NoStop}%
\bibitem [{\citenamefont {Graczyk}\ and\ \citenamefont
  {Kowal}(2020)}]{Graczyk:2019xwg}%
  \BibitemOpen
  \bibfield  {author} {\bibinfo {author} {\bibfnamefont {K.~M.}\ \bibnamefont
  {Graczyk}}\ and\ \bibinfo {author} {\bibfnamefont {B.~E.}\ \bibnamefont
  {Kowal}},\ }\href {\doibase 10.1103/PhysRevD.101.073002} {\bibfield
  {journal} {\bibinfo  {journal} {Phys. Rev. D}\ }\textbf {\bibinfo {volume}
  {101}},\ \bibinfo {pages} {073002} (\bibinfo {year} {2020})},\ \Eprint
  {http://arxiv.org/abs/1912.00064} {arXiv:1912.00064 [hep-ph]} \BibitemShut
  {NoStop}%
\bibitem [{\citenamefont {Graczyk}\ and\ \citenamefont
  {Kowal}(2019{\natexlab{a}})}]{Graczyk:2019opm}%
  \BibitemOpen
  \bibfield  {author} {\bibinfo {author} {\bibfnamefont {K.~M.}\ \bibnamefont
  {Graczyk}}\ and\ \bibinfo {author} {\bibfnamefont {B.~E.}\ \bibnamefont
  {Kowal}},\ }\href {\doibase 10.5506/APhysPolB.50.1771} {\bibfield  {journal}
  {\bibinfo  {journal} {Acta Phys. Polon. B}\ }\textbf {\bibinfo {volume}
  {50}},\ \bibinfo {pages} {1771} (\bibinfo {year}
  {2019}{\natexlab{a}})}\BibitemShut {NoStop}%
\bibitem [{\citenamefont {Fatima}\ \emph {et~al.}(2020)\citenamefont {Fatima},
  \citenamefont {Sajjad~Athar},\ and\ \citenamefont {Singh}}]{Fatima:2020pvv}%
  \BibitemOpen
  \bibfield  {author} {\bibinfo {author} {\bibfnamefont {A.}~\bibnamefont
  {Fatima}}, \bibinfo {author} {\bibfnamefont {M.}~\bibnamefont
  {Sajjad~Athar}}, \ and\ \bibinfo {author} {\bibfnamefont {S.~K.}\
  \bibnamefont {Singh}},\ }\href {\doibase 10.1103/PhysRevD.102.113009}
  {\bibfield  {journal} {\bibinfo  {journal} {Phys. Rev. D}\ }\textbf {\bibinfo
  {volume} {102}},\ \bibinfo {pages} {113009} (\bibinfo {year} {2020})},\
  \Eprint {http://arxiv.org/abs/2010.10311} {arXiv:2010.10311 [hep-ph]}
  \BibitemShut {NoStop}%
\bibitem [{\citenamefont {Tomalak}(2021)}]{Tomalak:2020zlv}%
  \BibitemOpen
  \bibfield  {author} {\bibinfo {author} {\bibfnamefont {O.}~\bibnamefont
  {Tomalak}},\ }\href {\doibase 10.1103/PhysRevD.103.013006} {\bibfield
  {journal} {\bibinfo  {journal} {Phys. Rev. D}\ }\textbf {\bibinfo {volume}
  {103}},\ \bibinfo {pages} {013006} (\bibinfo {year} {2021})},\ \Eprint
  {http://arxiv.org/abs/2008.03527} {arXiv:2008.03527 [hep-ph]} \BibitemShut
  {NoStop}%
\bibitem [{\citenamefont {Alam}\ \emph {et~al.}(2021)\citenamefont {Alam},
  \citenamefont {Alvarez-Ruso},\ and\ \citenamefont
  {Sato}}]{Luis_polarization}%
  \BibitemOpen
  \bibfield  {author} {\bibinfo {author} {\bibfnamefont {M.~R.}\ \bibnamefont
  {Alam}}, \bibinfo {author} {\bibfnamefont {L.}~\bibnamefont {Alvarez-Ruso}},
  \ and\ \bibinfo {author} {\bibfnamefont {T.}~\bibnamefont {Sato}},\ }\href
  {https://indico.phys.vt.edu/event/44/} {\enquote {\bibinfo {title} {Tau
  polarization in (anti-)neutrino-nucleon interactions},}\ } (\bibinfo {year}
  {2021}),\ \bibinfo {note} {new Directions in Neutrino-Nucleus
  Scattering}\BibitemShut {NoStop}%
\bibitem [{\citenamefont {Hagiwara}\ \emph {et~al.}(2005)\citenamefont
  {Hagiwara}, \citenamefont {Mawatari},\ and\ \citenamefont
  {Yokoya}}]{HAGIWARA2005140}%
  \BibitemOpen
  \bibfield  {author} {\bibinfo {author} {\bibfnamefont {K.}~\bibnamefont
  {Hagiwara}}, \bibinfo {author} {\bibfnamefont {K.}~\bibnamefont {Mawatari}},
  \ and\ \bibinfo {author} {\bibfnamefont {H.}~\bibnamefont {Yokoya}},\ }\href
  {\doibase https://doi.org/10.1016/j.nuclphysbps.2004.11.201} {\bibfield
  {journal} {\bibinfo  {journal} {Nuclear Physics B - Proceedings Supplements}\
  }\textbf {\bibinfo {volume} {139}},\ \bibinfo {pages} {140} (\bibinfo {year}
  {2005})},\ \bibinfo {note} {proceedings of the Third International Workshop
  on Neutrino-Nucleus Interactions in the Few-GeV Region}\BibitemShut {NoStop}%
\bibitem [{\citenamefont {Graczyk}\ and\ \citenamefont
  {Kowal}(2018)}]{Graczyk:2017rti}%
  \BibitemOpen
  \bibfield  {author} {\bibinfo {author} {\bibfnamefont {K.~M.}\ \bibnamefont
  {Graczyk}}\ and\ \bibinfo {author} {\bibfnamefont {B.~E.}\ \bibnamefont
  {Kowal}},\ }\href {\doibase 10.1103/PhysRevD.97.013001} {\bibfield  {journal}
  {\bibinfo  {journal} {Phys. Rev.}\ }\textbf {\bibinfo {volume} {D97}},\
  \bibinfo {pages} {013001} (\bibinfo {year} {2018})},\ \Eprint
  {http://arxiv.org/abs/1711.04868} {arXiv:1711.04868 [hep-ph]} \BibitemShut
  {NoStop}%
\bibitem [{\citenamefont {Graczyk}\ and\ \citenamefont
  {Kowal}(2019{\natexlab{b}})}]{Graczyk:2019blt}%
  \BibitemOpen
  \bibfield  {author} {\bibinfo {author} {\bibfnamefont {K.~M.}\ \bibnamefont
  {Graczyk}}\ and\ \bibinfo {author} {\bibfnamefont {B.~E.}\ \bibnamefont
  {Kowal}},\ }\href {\doibase 10.1103/PhysRevD.99.053002} {\bibfield  {journal}
  {\bibinfo  {journal} {Phys. Rev. D}\ }\textbf {\bibinfo {volume} {99}},\
  \bibinfo {pages} {053002} (\bibinfo {year} {2019}{\natexlab{b}})},\ \Eprint
  {http://arxiv.org/abs/1902.03671} {arXiv:1902.03671 [hep-ph]} \BibitemShut
  {NoStop}%
\bibitem [{\citenamefont {Thomas}\ and\ \citenamefont
  {Weise}(2010)}]{Thomas_book}%
  \BibitemOpen
  \bibfield  {author} {\bibinfo {author} {\bibfnamefont {A.}~\bibnamefont
  {Thomas}}\ and\ \bibinfo {author} {\bibfnamefont {W.}~\bibnamefont {Weise}},\
  }\href {http://books.google.pl/books?id=zbXvdYBmbJgC} {\emph {\bibinfo
  {title} {The Structure of the Nucleon}}}\ (\bibinfo  {publisher} {Wiley},\
  \bibinfo {year} {2010})\BibitemShut {NoStop}%
\bibitem [{\citenamefont {Hemmert}\ \emph {et~al.}(1995)\citenamefont
  {Hemmert}, \citenamefont {Holstein},\ and\ \citenamefont
  {Mukhopadhyay}}]{Hemmert:1994ky}%
  \BibitemOpen
  \bibfield  {author} {\bibinfo {author} {\bibfnamefont {T.~R.}\ \bibnamefont
  {Hemmert}}, \bibinfo {author} {\bibfnamefont {B.~R.}\ \bibnamefont
  {Holstein}}, \ and\ \bibinfo {author} {\bibfnamefont {N.~C.}\ \bibnamefont
  {Mukhopadhyay}},\ }\href {\doibase 10.1103/PhysRevD.51.158} {\bibfield
  {journal} {\bibinfo  {journal} {Phys. Rev. D}\ }\textbf {\bibinfo {volume}
  {51}},\ \bibinfo {pages} {158} (\bibinfo {year} {1995})},\ \Eprint
  {http://arxiv.org/abs/hep-ph/9409323} {arXiv:hep-ph/9409323} \BibitemShut
  {NoStop}%
\bibitem [{\citenamefont {Barquilla-Cano}\ \emph {et~al.}(2007)\citenamefont
  {Barquilla-Cano}, \citenamefont {Buchmann},\ and\ \citenamefont
  {Hernandez}}]{BarquillaCano:2007yk}%
  \BibitemOpen
  \bibfield  {author} {\bibinfo {author} {\bibfnamefont {D.}~\bibnamefont
  {Barquilla-Cano}}, \bibinfo {author} {\bibfnamefont {A.~J.}\ \bibnamefont
  {Buchmann}}, \ and\ \bibinfo {author} {\bibfnamefont {E.}~\bibnamefont
  {Hernandez}},\ }\href {\doibase 10.1103/PhysRevC.77.019903} {\bibfield
  {journal} {\bibinfo  {journal} {Phys. Rev. C}\ }\textbf {\bibinfo {volume}
  {75}},\ \bibinfo {pages} {065203} (\bibinfo {year} {2007})},\ \bibinfo {note}
  {[Erratum: Phys.Rev.C 77, 019903 (2008)]},\ \Eprint
  {http://arxiv.org/abs/0705.3297} {arXiv:0705.3297 [nucl-th]} \BibitemShut
  {NoStop}%
\bibitem [{\citenamefont {Geng}\ \emph {et~al.}(2008)\citenamefont {Geng},
  \citenamefont {Martin~Camalich}, \citenamefont {Alvarez-Ruso},\ and\
  \citenamefont {Vicente~Vacas}}]{Geng:2008bm}%
  \BibitemOpen
  \bibfield  {author} {\bibinfo {author} {\bibfnamefont {L.~S.}\ \bibnamefont
  {Geng}}, \bibinfo {author} {\bibfnamefont {J.}~\bibnamefont
  {Martin~Camalich}}, \bibinfo {author} {\bibfnamefont {L.}~\bibnamefont
  {Alvarez-Ruso}}, \ and\ \bibinfo {author} {\bibfnamefont {M.~J.}\
  \bibnamefont {Vicente~Vacas}},\ }\href {\doibase 10.1103/PhysRevD.78.014011}
  {\bibfield  {journal} {\bibinfo  {journal} {Phys. Rev. D}\ }\textbf {\bibinfo
  {volume} {78}},\ \bibinfo {pages} {014011} (\bibinfo {year} {2008})},\
  \Eprint {http://arxiv.org/abs/0801.4495} {arXiv:0801.4495 [hep-ph]}
  \BibitemShut {NoStop}%
\bibitem [{\citenamefont {Procura}(2008)}]{Procura:2008ze}%
  \BibitemOpen
  \bibfield  {author} {\bibinfo {author} {\bibfnamefont {M.}~\bibnamefont
  {Procura}},\ }\href {\doibase 10.1103/PhysRevD.78.094021} {\bibfield
  {journal} {\bibinfo  {journal} {Phys. Rev. D}\ }\textbf {\bibinfo {volume}
  {78}},\ \bibinfo {pages} {094021} (\bibinfo {year} {2008})},\ \Eprint
  {http://arxiv.org/abs/0803.4291} {arXiv:0803.4291 [hep-ph]} \BibitemShut
  {NoStop}%
\bibitem [{\citenamefont {\"Unal}\ \emph {et~al.}(2019)\citenamefont {\"Unal},
  \citenamefont {K\"u\c{c}\"ukarslan},\ and\ \citenamefont
  {Scherer}}]{Unal:2018ruo}%
  \BibitemOpen
  \bibfield  {author} {\bibinfo {author} {\bibfnamefont {Y.}~\bibnamefont
  {\"Unal}}, \bibinfo {author} {\bibfnamefont {A.}~\bibnamefont
  {K\"u\c{c}\"ukarslan}}, \ and\ \bibinfo {author} {\bibfnamefont
  {S.}~\bibnamefont {Scherer}},\ }\href {\doibase 10.1103/PhysRevD.99.014012}
  {\bibfield  {journal} {\bibinfo  {journal} {Phys. Rev. D}\ }\textbf {\bibinfo
  {volume} {99}},\ \bibinfo {pages} {014012} (\bibinfo {year} {2019})},\
  \Eprint {http://arxiv.org/abs/1808.03046} {arXiv:1808.03046 [hep-ph]}
  \BibitemShut {NoStop}%
\bibitem [{\citenamefont {Kucukarslan}\ \emph {et~al.}(2014)\citenamefont
  {Kucukarslan}, \citenamefont {Ozdem},\ and\ \citenamefont
  {Ozpineci}}]{Kucukarslan:2014bla}%
  \BibitemOpen
  \bibfield  {author} {\bibinfo {author} {\bibfnamefont {A.}~\bibnamefont
  {Kucukarslan}}, \bibinfo {author} {\bibfnamefont {U.}~\bibnamefont {Ozdem}},
  \ and\ \bibinfo {author} {\bibfnamefont {A.}~\bibnamefont {Ozpineci}},\
  }\href {\doibase 10.1103/PhysRevD.90.054002} {\bibfield  {journal} {\bibinfo
  {journal} {Phys. Rev. D}\ }\textbf {\bibinfo {volume} {90}},\ \bibinfo
  {pages} {054002} (\bibinfo {year} {2014})},\ \Eprint
  {http://arxiv.org/abs/1407.3735} {arXiv:1407.3735 [hep-ph]} \BibitemShut
  {NoStop}%
\bibitem [{\citenamefont {Alexandrou}\ \emph {et~al.}(2007)\citenamefont
  {Alexandrou}, \citenamefont {Koutsou}, \citenamefont {Leontiou},
  \citenamefont {Negele},\ and\ \citenamefont {Tsapalis}}]{Alexandrou:2009vqd}%
  \BibitemOpen
  \bibfield  {author} {\bibinfo {author} {\bibfnamefont {C.}~\bibnamefont
  {Alexandrou}}, \bibinfo {author} {\bibfnamefont {G.}~\bibnamefont {Koutsou}},
  \bibinfo {author} {\bibfnamefont {T.}~\bibnamefont {Leontiou}}, \bibinfo
  {author} {\bibfnamefont {J.~W.}\ \bibnamefont {Negele}}, \ and\ \bibinfo
  {author} {\bibfnamefont {A.}~\bibnamefont {Tsapalis}},\ }\href {\doibase
  10.1103/PhysRevD.80.099901} {\bibfield  {journal} {\bibinfo  {journal} {Phys.
  Rev. D}\ }\textbf {\bibinfo {volume} {76}},\ \bibinfo {pages} {094511}
  (\bibinfo {year} {2007})},\ \bibinfo {note} {[Erratum: Phys.Rev.D 80, 099901
  (2009)]},\ \Eprint {http://arxiv.org/abs/0706.3011} {arXiv:0706.3011
  [hep-lat]} \BibitemShut {NoStop}%
\bibitem [{\citenamefont {Alexandrou}\ \emph {et~al.}(2013)\citenamefont
  {Alexandrou}, \citenamefont {Gregory}, \citenamefont {Korzec}, \citenamefont
  {Koutsou}, \citenamefont {Negele}, \citenamefont {Sato},\ and\ \citenamefont
  {Tsapalis}}]{Alexandrou:2013opa}%
  \BibitemOpen
  \bibfield  {author} {\bibinfo {author} {\bibfnamefont {C.}~\bibnamefont
  {Alexandrou}}, \bibinfo {author} {\bibfnamefont {E.~B.}\ \bibnamefont
  {Gregory}}, \bibinfo {author} {\bibfnamefont {T.}~\bibnamefont {Korzec}},
  \bibinfo {author} {\bibfnamefont {G.}~\bibnamefont {Koutsou}}, \bibinfo
  {author} {\bibfnamefont {J.~W.}\ \bibnamefont {Negele}}, \bibinfo {author}
  {\bibfnamefont {T.}~\bibnamefont {Sato}}, \ and\ \bibinfo {author}
  {\bibfnamefont {A.}~\bibnamefont {Tsapalis}},\ }\href {\doibase
  10.1103/PhysRevD.87.114513} {\bibfield  {journal} {\bibinfo  {journal} {Phys.
  Rev. D}\ }\textbf {\bibinfo {volume} {87}},\ \bibinfo {pages} {114513}
  (\bibinfo {year} {2013})},\ \Eprint {http://arxiv.org/abs/1304.4614}
  {arXiv:1304.4614 [hep-lat]} \BibitemShut {NoStop}%
\bibitem [{\citenamefont {Liu}\ \emph {et~al.}(1995)\citenamefont {Liu},
  \citenamefont {Mukhopadhyay},\ and\ \citenamefont {Zhang}}]{Liu:1995bu}%
  \BibitemOpen
  \bibfield  {author} {\bibinfo {author} {\bibfnamefont {J.}~\bibnamefont
  {Liu}}, \bibinfo {author} {\bibfnamefont {N.~C.}\ \bibnamefont
  {Mukhopadhyay}}, \ and\ \bibinfo {author} {\bibfnamefont {L.-s.}\
  \bibnamefont {Zhang}},\ }\href {\doibase 10.1103/PhysRevC.52.1630} {\bibfield
   {journal} {\bibinfo  {journal} {Phys. Rev. C}\ }\textbf {\bibinfo {volume}
  {52}},\ \bibinfo {pages} {1630} (\bibinfo {year} {1995})},\ \Eprint
  {http://arxiv.org/abs/hep-ph/9506389} {arXiv:hep-ph/9506389} \BibitemShut
  {NoStop}%
\bibitem [{\citenamefont {Lalakulich}\ and\ \citenamefont
  {Paschos}(2005)}]{Lalakulich:2005cs}%
  \BibitemOpen
  \bibfield  {author} {\bibinfo {author} {\bibfnamefont {O.}~\bibnamefont
  {Lalakulich}}\ and\ \bibinfo {author} {\bibfnamefont {E.~A.}\ \bibnamefont
  {Paschos}},\ }\href {\doibase 10.1103/PhysRevD.71.074003} {\bibfield
  {journal} {\bibinfo  {journal} {Phys. Rev.}\ }\textbf {\bibinfo {volume}
  {D71}},\ \bibinfo {pages} {074003} (\bibinfo {year} {2005})},\ \Eprint
  {http://arxiv.org/abs/hep-ph/0501109} {arXiv:hep-ph/0501109 [hep-ph]}
  \BibitemShut {NoStop}%
\bibitem [{\citenamefont {Alvarez-Ruso}\ \emph {et~al.}(1999)\citenamefont
  {Alvarez-Ruso}, \citenamefont {Singh},\ and\ \citenamefont
  {Vicente~Vacas}}]{AlvarezRuso:1998hi}%
  \BibitemOpen
  \bibfield  {author} {\bibinfo {author} {\bibfnamefont {L.}~\bibnamefont
  {Alvarez-Ruso}}, \bibinfo {author} {\bibfnamefont {S.~K.}\ \bibnamefont
  {Singh}}, \ and\ \bibinfo {author} {\bibfnamefont {M.~J.}\ \bibnamefont
  {Vicente~Vacas}},\ }\href {\doibase 10.1103/PhysRevC.59.3386} {\bibfield
  {journal} {\bibinfo  {journal} {Phys. Rev. C}\ }\textbf {\bibinfo {volume}
  {59}},\ \bibinfo {pages} {3386} (\bibinfo {year} {1999})},\ \Eprint
  {http://arxiv.org/abs/nucl-th/9804007} {arXiv:nucl-th/9804007} \BibitemShut
  {NoStop}%
\bibitem [{\citenamefont {Leitner}\ \emph {et~al.}(2008)\citenamefont
  {Leitner}, \citenamefont {Buss}, \citenamefont {Mosel},\ and\ \citenamefont
  {Alvarez-Ruso}}]{Leitner:2008fg}%
  \BibitemOpen
  \bibfield  {author} {\bibinfo {author} {\bibfnamefont {T.}~\bibnamefont
  {Leitner}}, \bibinfo {author} {\bibfnamefont {O.}~\bibnamefont {Buss}},
  \bibinfo {author} {\bibfnamefont {U.}~\bibnamefont {Mosel}}, \ and\ \bibinfo
  {author} {\bibfnamefont {L.}~\bibnamefont {Alvarez-Ruso}},\ }\href {\doibase
  10.22323/1.074.0009} {\bibfield  {journal} {\bibinfo  {journal} {PoS}\
  }\textbf {\bibinfo {volume} {NUFACT08}},\ \bibinfo {pages} {009} (\bibinfo
  {year} {2008})},\ \Eprint {http://arxiv.org/abs/0809.3986} {arXiv:0809.3986
  [nucl-th]} \BibitemShut {NoStop}%
\bibitem [{\citenamefont {de~Swart}\ \emph {et~al.}(1997)\citenamefont
  {de~Swart}, \citenamefont {Rentmeester},\ and\ \citenamefont
  {Timmermans}}]{deSwart:1997ep}%
  \BibitemOpen
  \bibfield  {author} {\bibinfo {author} {\bibfnamefont {J.~J.}\ \bibnamefont
  {de~Swart}}, \bibinfo {author} {\bibfnamefont {M.~C.~M.}\ \bibnamefont
  {Rentmeester}}, \ and\ \bibinfo {author} {\bibfnamefont {R.~G.~E.}\
  \bibnamefont {Timmermans}},\ }\bibfield  {booktitle} {\emph {\bibinfo
  {booktitle} {{Meson nucleon physics and the structure of the nucleon.
  Proceedings, 7th International Symposium, MENU'97, Vancouver, Canada, July
  28-August 1, 1997}}},\ }\href@noop {} {\bibfield  {journal} {\bibinfo
  {journal} {PiN Newslett.}\ }\textbf {\bibinfo {volume} {13}},\ \bibinfo
  {pages} {96} (\bibinfo {year} {1997})},\ \Eprint
  {http://arxiv.org/abs/nucl-th/9802084} {arXiv:nucl-th/9802084 [nucl-th]}
  \BibitemShut {NoStop}%
\bibitem [{\citenamefont {Rarita}\ and\ \citenamefont
  {Schwinger}(1941)}]{Rarita:1941mf}%
  \BibitemOpen
  \bibfield  {author} {\bibinfo {author} {\bibfnamefont {W.}~\bibnamefont
  {Rarita}}\ and\ \bibinfo {author} {\bibfnamefont {J.}~\bibnamefont
  {Schwinger}},\ }\href {\doibase 10.1103/PhysRev.60.61} {\bibfield  {journal}
  {\bibinfo  {journal} {Phys. Rev.}\ }\textbf {\bibinfo {volume} {60}},\
  \bibinfo {pages} {61} (\bibinfo {year} {1941})}\BibitemShut {NoStop}%
\bibitem [{\citenamefont {Jones}\ and\ \citenamefont
  {Scadron}(1973)}]{Jones:1972ky}%
  \BibitemOpen
  \bibfield  {author} {\bibinfo {author} {\bibfnamefont {H.}~\bibnamefont
  {Jones}}\ and\ \bibinfo {author} {\bibfnamefont {M.}~\bibnamefont
  {Scadron}},\ }\href {\doibase 10.1016/0003-4916(73)90476-4} {\bibfield
  {journal} {\bibinfo  {journal} {Annals Phys.}\ }\textbf {\bibinfo {volume}
  {81}},\ \bibinfo {pages} {1} (\bibinfo {year} {1973})}\BibitemShut {NoStop}%
\bibitem [{\citenamefont {Barish}\ \emph {et~al.}(1979)\citenamefont {Barish}
  \emph {et~al.}}]{Barish:1978pj}%
  \BibitemOpen
  \bibfield  {author} {\bibinfo {author} {\bibfnamefont {S.~J.}\ \bibnamefont
  {Barish}} \emph {et~al.},\ }\href {\doibase 10.1103/PhysRevD.19.2521}
  {\bibfield  {journal} {\bibinfo  {journal} {Phys. Rev.}\ }\textbf {\bibinfo
  {volume} {D19}},\ \bibinfo {pages} {2521} (\bibinfo {year}
  {1979})}\BibitemShut {NoStop}%
\bibitem [{\citenamefont {Vermaseren}(2000)}]{Vermaseren:2000nd}%
  \BibitemOpen
  \bibfield  {author} {\bibinfo {author} {\bibfnamefont {J.~A.~M.}\
  \bibnamefont {Vermaseren}},\ }\href@noop {} {\  (\bibinfo {year} {2000})},\
  \Eprint {http://arxiv.org/abs/math-ph/0010025} {arXiv:math-ph/0010025
  [math-ph]} \BibitemShut {NoStop}%
\bibitem [{\citenamefont {Mertig}\ \emph {et~al.}(1991)\citenamefont {Mertig},
  \citenamefont {Bohm},\ and\ \citenamefont {Denner}}]{Mertig:1990an}%
  \BibitemOpen
  \bibfield  {author} {\bibinfo {author} {\bibfnamefont {R.}~\bibnamefont
  {Mertig}}, \bibinfo {author} {\bibfnamefont {M.}~\bibnamefont {Bohm}}, \ and\
  \bibinfo {author} {\bibfnamefont {A.}~\bibnamefont {Denner}},\ }\href
  {\doibase 10.1016/0010-4655(91)90130-D} {\bibfield  {journal} {\bibinfo
  {journal} {Comput.Phys.Commun.}\ }\textbf {\bibinfo {volume} {64}},\ \bibinfo
  {pages} {345} (\bibinfo {year} {1991})}\BibitemShut {NoStop}%
\bibitem [{\citenamefont {Shtabovenko}\ \emph {et~al.}(2016)\citenamefont
  {Shtabovenko}, \citenamefont {Mertig},\ and\ \citenamefont
  {Orellana}}]{Shtabovenko:2016sxi}%
  \BibitemOpen
  \bibfield  {author} {\bibinfo {author} {\bibfnamefont {V.}~\bibnamefont
  {Shtabovenko}}, \bibinfo {author} {\bibfnamefont {R.}~\bibnamefont {Mertig}},
  \ and\ \bibinfo {author} {\bibfnamefont {F.}~\bibnamefont {Orellana}},\
  }\href {\doibase 10.1016/j.cpc.2016.06.008} {\bibfield  {journal} {\bibinfo
  {journal} {Comput. Phys. Commun.}\ }\textbf {\bibinfo {volume} {207}},\
  \bibinfo {pages} {432} (\bibinfo {year} {2016})}\BibitemShut {NoStop}%
\end{thebibliography}%

\end{document}